\documentclass[12pt]{article}

\usepackage{graphicx}
\usepackage{multirow}%multirow
\usepackage{booktabs}
\usepackage{array}
\usepackage{booktabs,subcaption,amsfonts,dcolumn}
\usepackage{soul}
\usepackage{authblk} % Manages authors' affiliation
\usepackage{natbib} %comment out if you do not have the package
\usepackage{url} % not crucial - just used below for the URL 
\usepackage{amsmath}
\usepackage{flafter}           % floats never appear before their position in the text
\usepackage[section]{placeins} % floats never cross the start of the next section
\usepackage{mathtools}

\newcommand{\blind}{0}
\usepackage{enumitem}
\usepackage[doublespacing]{setspace}
\usepackage{bm}
\usepackage{amssymb}
\usepackage[font=scriptsize]{caption} % To reduce caption font size
\usepackage[ruled,vlined]{algorithm2e}

\DeclareMathOperator{\Cov}{Cov}
\DeclareMathOperator{\E}{E}

\DeclareMathOperator{\diag}{diag}
  
\newcolumntype{M}[1]{>{\centering\arraybackslash}m{#1}}

\usepackage{xcolor}

\usepackage{booktabs}

\usepackage{pifont} % For checkmarks

\begin{document}

\def\spacingset#1{\renewcommand{\baselinestretch}%
{#1}\small\normalsize} \spacingset{1}

%%%%%%%%%%%%%%%%%%%%%%%%%%%%%%%%%%%%%%%%%%%%%%%%%%%%%%%%%%%%%%%%%%%%%%%%%%%%%%

\if0\blind
{
\title{ \textbf{Statistical Process Monitoring of Functional Linear Models}}

\author[1]{Davide Forcina}
\author[1]{Antonio Lepore
\thanks{Corresponding author. e-mail: \texttt{antonio.lepore@unina.it}}}
\author[2]{Kamran Paynabar}

\affil[1]{Department of Industrial Engineering, University of Naples Federico II, Naples, Italy}
\affil[2]{School of Industrial and Systems Engineering, Georgia Institute of Technology, Atlanta, GA}

\setcounter{Maxaffil}{0}
\renewcommand\Affilfont{\itshape\small}
\date{}
\maketitle

} \fi

\if1\blind
{
  \bigskip
  \bigskip
  \bigskip
  \begin{center}
    {\LARGE\bf Statistical Process Monitoring of Functional Linear Models}
\end{center}
  \bigskip
  \bigskip
  \bigskip
} \fi

\begin{abstract}
\noindent Monitoring the stability over time of the relationship between a functional response and multiple functional covariates is important in many industrial applications, either because the relationship itself is the quality characteristic of interest or because it underpins a predictive model whose validity must be verified as the process evolves. 
Existing methods either assume the model estimated in Phase I to be stable and monitor
the resulting residuals, or re-estimate its coefficients at each sampling stage, which
requires several observations to be available at a time.
In this article, we propose the regression coefficient function (RCF) control chart to monitor functional linear models (FLMs) in which, at each sampling stage, an individual observation of the functional response and an arbitrary number of functional covariates become available. 
The proposed method employs an exponentially weighted moving loss to sequentially update estimates of the regression coefficient functions and the intercept function, which are monitored through Hotelling's $T^2$ and squared prediction error ($SPE$) control charts. 
A post-signal diagnostic procedure is also provided to support the identification of the source of the detected change. 
A Monte Carlo simulation study demonstrates that the proposed method outperforms
state-of-the-art methods, particularly against small and persistent shifts, and a case study on monitoring the relationship between the air/fuel ratio and engine state variables in an automotive after-treatment system illustrates its practical applicability.
\end{abstract}

\noindent%

\spacingset{1.45} % DON'T change the spacing!
% \spacingset{2} % DON'T change the spacing!

\section{Introduction}
\label{sec_intro}
In many industrial scenarios, the process state is best represented by the relationship between the quality characteristic of interest, which can be scalar, multivariate, or functional, and one or more concurrent process variables rather than in the quality characteristic alone. Changes in the relationship may in fact directly invalidate the predictive model built on it, e.g., for forecasting \citep{brown1975techniques} or statistical process monitoring (SPM) purposes.

An example is the case of modern diesel engines, such as the one considered in Section~\ref{sec_real}, that aim to reduce nitrogen-oxide ($\mathrm{NO_x}$) emissions through a $\mathrm{NO_x}$ storage catalyst (NSC), which is periodically regenerated \citep{pacella2018unsupervised}. The effectiveness of the regeneration is measured by the relative air/fuel ratio signal, also known as the $\lambda$ signal, collected from on-board sensors along with many other engine state variables.
\cite{gahrooei2019process} highlight, from a systems engineering perspective, the advantages of a model that predicts the $\lambda$ signal from engine state variables to adjust and control engine operating conditions and identify faulty sensors, whereas \cite{gomez2020functional} stress the importance of examining this relationship to diagnose root causes of system faults that cannot be identified from the $\lambda$ signal alone.

In their seminal work, \cite{kang2000line} considered the monitoring of a process whose state is characterized by a linear relationship rather than by one or more scalar quality characteristics, assuming a sample of observations to be available at each sampling stage, a problem that later became known as profile monitoring \citep{woodall2004using,noorossana2011statistical}.
As is customary in SPM, monitoring occurs in two phases \citep{qiu2013introduction,montgomery2020introduction}. In Phase I, a clean set of observations, assumed to be representative of the in-control (IC) state of the process and hereinafter referred to as the Phase I sample, is identified and used to characterize this state; in Phase II, new observations are monitored to detect a transition to an out-of-control (OC) state.
To address the profile monitoring problem, \cite{kang2000line} essentially proposed two strategies.
One, hereinafter referred to as the \textit{indirect} approach,  compares the incoming sample of observations at each sampling stage with a reference model estimated in Phase I and monitors the resulting deviations. The other, hereinafter referred to as the \textit{direct} approach, re-estimates the model coefficients using the incoming sample of observations and tests them directly.
In this way, the direct approach can naturally attribute an alarm signal to a specific component, and the updated estimate can describe how the relationship has changed, e.g., to support model revision. 
Diagnostics under the indirect approach can instead only locate the change in the response domain rather than in the regression coefficients.

Unfortunately, unlike the indirect approach, the direct approach cannot be applied to individual observations with standard estimation methods, because a single observation contains insufficient information to obtain the model coefficient estimates to be monitored at each sampling stage.
To address this issue, the starting point is to move beyond the SPM literature, where \cite{farley1970test} developed a locally most powerful test for a shifting slope coefficient in a scalar linear model. \cite{brown1975techniques} proposed tests based on recursive residuals to check whether regression relationships remain constant over time, while \cite{kim1989likelihood} considered likelihood ratio tests to detect a change-point in simple linear regression. 
\cite{horvath2004monitoring} later provided a formal asymptotic framework for sequential monitoring, proposing two cumulative sum (CUSUM) schemes on residuals, applicable after a Phase I sample is available, to detect a structural change in a linear model.
These contributions test for a change in the regression coefficients but are not designed as control charts, whose performance is usually measured by the average run length ($ARL$), that is, the average number of observations required to signal. In particular, their critical values control a type I error probability, either on a fixed sample or over the whole monitoring horizon, rather than a prespecified IC $ARL$, denoted by $ARL_0$.

In the functional setting, the estimation of a functional linear model (FLM) is a known ill-posed inverse problem even for large samples, owing to the infinite dimensionality of the coefficient functions, and has been extensively studied for scalar-on-function \citep{ramsay1991some,cardot2003spline,yao2005functional}, function-on-scalar \citep{ramsay}, and function-on-function relationships \citep{matsui2009regularized,luo2017function,gahrooei2019process}.
When only individual functional observations are available at each acquisition in Phase II, the coefficient functions cannot be reliably re-estimated from a single observation, which precludes a direct approach unless sequential methods are used.
The only such method is that of \cite{aue2014dependent}, who extend sequential monitoring to a function-on-function linear relationship. To address the infinite dimensionality, the authors apply functional principal component analysis to both the covariate and response functions, and monitor using a CUSUM-type procedure.
The resulting detector is compared with a curved threshold function, whose critical values come from the asymptotic null distribution of a functional of independent Brownian motions and depend on the number of retained components.
However, their procedure is limited to a single functional covariate and overlooks variation outside the retained components, so changes in those directions go undetected.
Retaining only a few components from Phase I can cause the monitoring statistics to fail to span the subspace in which OC states may arise \citep{woodall2004using}.
Moreover, the monitoring procedure in \cite{aue2014dependent} is not natively designed to achieve a prespecified $ARL_0$ and cannot provide post-signal diagnostics to guide identification of the possible source of the detected change.

In this paper, we aim to overcome these limitations and develop a new control chart, the regression coefficient function (RCF) control chart, for direct SPM of an FLM when the functional response depends on multiple functional covariates and only individual functional observations are available at each Phase II sampling stage.
Beyond the automotive application described above, motivating problems arise in environmental monitoring \citep{aue2014dependent}, manufacturing process monitoring \citep{capezza2026functional}, and emissions monitoring \citep{centofanti2021bfunctional}.
Within the taxonomy outlined above, the RCF control chart represents the functional counterpart of the direct approach for individual observations, whereas the functional regression control chart (FRCC) of \cite{centofanti2021bfunctional} is the functional counterpart of the indirect approach. The latter assumes a stable FLM estimated from the Phase I sample and monitors the functional residuals of each individual observation through Hotelling's $T^2$ and squared prediction error ($SPE$) control charts.

The proposed RCF control chart makes three main methodological contributions. First, it addresses the ill-posed single-observation estimation problem in FLMs with multiple covariates through an exponentially weighted moving loss \citep{luxenberg2024exponentially}. At each sampling stage, it accumulates information across observations while progressively discounting past ones, updating estimates of the \textit{stacked coefficient vector}, defined as the vector containing the intercept and regression coefficient functions. 
The memory-type nature of the proposed strategy enables detection of small, persistent changes that a method based only on the current observation might not reveal.
Second, the joint use of Hotelling's $T^2$ and  $SPE$ control charts for the stacked coefficient vector allows monitoring variation both within the retained principal component subspace and in its orthogonal residual subspace. Third, we provide a post-signal diagnostic procedure to identify which component of the stacked coefficient vector drives a detected change. An extensive Monte Carlo simulation study compares the proposed method with state-of-the-art methods, and a case study on monitoring the functional relationship between the $\lambda$ signal and engine state variables in an automotive application illustrates its practical applicability.

The remainder of the article is organized as follows. Section \ref{methodology} fully describes the proposed RCF control chart. Section \ref{sec_sim} compares the proposed method with state-of-the-art methods using a Monte Carlo simulation study. Section \ref{sec_real} presents the case study. Section \ref{sec_conclusions} concludes the article. The Supplementary Materials include additional details on data generation in the simulation study, additional simulation results for the post-signal diagnostic procedure, and sensitivity analyses of tuning parameters. All calculations and plots were obtained using the programming language \textsf{R} \citep{r2021}.

\section{Methodology}
\label{methodology}
This section describes the proposed RCF control chart in five subsections. Section \ref{setting} introduces the statistical setting, 
defines the functional linear model, and formulates the monitoring problem as a hypothesis test on the stability of the regression coefficient vector. Section \ref{seqest} presents the sequential estimation procedure, which updates the regression coefficient functions at each sampling stage via an exponentially weighted moving loss. Section~\ref{monitoring} describes the monitoring procedure, where the updated coefficient estimates are projected onto a low-dimensional space via Multivariate Functional Principal Component Analysis (MFPCA) and monitored through $T^2$ and $SPE$ control chart statistics. Section \ref{post-signal} introduces the post-signal diagnostic procedure to identify the regression coefficient functions that may have shifted off target. Section \ref{implementation} provides implementation details.

\subsection{The Regression Coefficient Function Control Chart Setting}
\label{setting}
Let $L^2(\mathcal{D})$ denote the Hilbert space of square-integrable functions defined in the compact set $\mathcal{D} \subseteq \mathbb{R}$, with the inner product $\langle f,g\rangle_\mathcal{D} = \int_{\mathcal{D}}f(s)g(s)ds$ and the norm $\|\cdot\| = \sqrt{\langle\cdot,\cdot\rangle_\mathcal{D}}$.
Let $Y(t)$ denote the functional response variable whose realizations are in $L^2(\mathcal{T})$ and let $\boldsymbol{X}(s) = \left(X_1(s),\dots,X_p(s)\right)^T$ be the vector of functional covariates whose realizations are in $\left(L^2(\mathcal{S})\right)^p$, that is, the Hilbert space $\mathbb{H}$ whose elements are vectors of functions in $L^2(\mathcal{S})$. The inner product of two function vectors $\boldsymbol{f} = \left(f_1,\dots,f_p\right)^T$ and $\boldsymbol{g} = \left(g_1,\dots,g_p\right)^T$ in $\mathbb{H}$ is 
$\langle\boldsymbol{f},\boldsymbol{g}\rangle_{\mathbb{H}}= \sum_{j = 1}^p\langle f_j,g_j \rangle_\mathcal{S}$ and the norm $\|\cdot\|_\mathbb{H} = \sqrt{\langle\cdot,\cdot\rangle_{\mathbb{H}}}$.
Let $\{(Y_i, \boldsymbol{X}_i)\}_{i=1}^{\infty}$ denote a sequence of independent functional observations, 
where $\boldsymbol{X}_i(s) = \left(X_{1i}(s), \dots, X_{pi}(s)\right)^T \in \mathbb{H}$ 
and $Y_i \in L^2(\mathcal{T})$ are the vector of functional covariates and the functional 
response at sampling stage $i$, respectively.
We assume that the functional covariates linearly influence the functional response through the following FLM
\begin{equation}
\label{flm}
    Y_i(t) =  \beta_{0i}(t) + \sum_{j = 1}^p\int_\mathcal{S}\beta_{ji}(s,t)X_{ji}(s)ds + \varepsilon_i(t), \quad t \in \mathcal{T}, \quad i = 1,2,\dots,
\end{equation}
where the intercept function $\beta_{0i}(t)$ is in $L^2(\mathcal{T})$, the regression coefficient vector $\left(\beta_{1i},\dots,\beta_{pi} \right)^T$ is in $\left(L^2(\mathcal{S}\times\mathcal{T})\right)^p$, and the random error function $\varepsilon_i$ has $\E(\varepsilon_i) = 0$, and is independent of $\boldsymbol{X}_i$, for $i = 1,2,\dots$.

The regression coefficient vector 
$\left(\beta_{1i}, \dots, \beta_{pi}\right)^T$ expresses the 
functional relationship between $Y_i$ and $\boldsymbol{X}_i$ at sampling stage $i$, and together with the intercept function $\beta_{0i}$ is the quantity to be monitored.
In particular, let $\boldsymbol{\beta}_i = \left(\beta_{0i},\beta_{1i}, \dots, \beta_{pi} \right)^T$ be the vector obtained by stacking the intercept function and the vector of regression coefficient functions.
It belongs to the product Hilbert space $\mathbb{H}_\beta = L^2(\mathcal{T}) \times \left(L^2(\mathcal{S} \times 
\mathcal{T})\right)^p$, whose elements are vectors $\boldsymbol{f} = 
\left(f_0, f_1, \dots, f_p\right)^T$ with $f_0 \in L^2(\mathcal{T})$ and 
$f_j \in L^2(\mathcal{S} \times \mathcal{T})$, $j = 1,\dots,p$. The inner product of two such vectors $\boldsymbol{f}$ and $\boldsymbol{g}$ is 
$\langle\boldsymbol{f}, \boldsymbol{g}\rangle_{\mathbb{H}_\beta} = \sum_{j=0}^p \langle f_j, g_j\rangle_{\mathcal{D}_j}$, where $\mathcal{D}_0 = \mathcal{T}$, $\mathcal{D}_j = \mathcal{S} \times \mathcal{T}$ for $j \geq 1$, with the associated norm $\|\cdot\|_{\mathbb{H}_\beta} = \sqrt{\langle\cdot,\cdot\rangle_{\mathbb{H}_\beta}}$.

The control chart we want to introduce for the monitoring of functional linear regression relationships is based on the following null (the process is IC) and alternative (the process is OC) hypotheses
\begin{equation}
\label{hyp}
    \begin{cases}
        H_0: \boldsymbol{\beta}_i = \boldsymbol{\beta}_0, \quad i = 1, 2, \dots, \\[6pt]
        H_1: \exists\, i^* \text{ such that } \boldsymbol{\beta}_i = \boldsymbol{\beta}_0 
        \text{ for all } i < i^* \text{ and } \boldsymbol{\beta}_i \neq 
        \boldsymbol{\beta}_0 \text{ for all } i \geq i^*,
    \end{cases}
\end{equation}
where $\boldsymbol{\beta}_0 = \left(\beta_{00}, \beta_{10},\dots,\beta_{p0}\right)^T$ is the IC vector of parameters.
The IC coefficient vector $\boldsymbol{\beta}_0$ may be known, e.g., from prior physical or engineering knowledge.
When the IC coefficient vector is unknown, we estimate it from the Phase I sample using the procedure of \cite{gahrooei2019process}, who adopt a penalized batch procedure through a LASSO penalty on the centered versions of the functional response and functional covariates.
This implies the IC intercept function $\beta_{00}(t)$ is equal to 0 in Equation \eqref{hyp}. Therefore, a shift in the mean level of the functional response would cause the sequentially estimated intercept function to depart from zero.
We omit additional details here and refer the reader to that work for a complete description.

\subsection{Sequential Estimation}
\label{seqest}
To test the hypotheses in Equation \eqref{hyp} at each sampling stage $i = 1,2,\dots$, an estimate of the stacked coefficient vector $\boldsymbol{\beta}_i$ must be obtained from the current functional observation $\left(Y_i,\boldsymbol{X}_i\right)$, which is supposed to be observed on discrete grids of size $n$ for the functional response and size $m_j$, $j=1,\dots,p$, for the functional covariates, respectively. It is also assumed that the grid size for the response and each of the covariates does not change from sample to sample.
To estimate the model parameters $\beta_{0i}(t)$ and $\beta_{ji}(s,t)$, $j = 1,\dots,p$, \cite{gahrooei2019process} proposed a penalized estimation procedure based on the expansion of the regression coefficient functions $\beta_{j}(s,t)$, $j = 1,\dots,p$, when a sample of independent realizations $\left\{Y_i(t),\boldsymbol{X}_i(s)\right\}_{i = 1,\dots,m}$ of size $m$ is available, without considering the intercept function $\beta_0(t)$.
In what follows, we present the estimation procedure for a single observation $\left(Y_i, \boldsymbol{X}_i\right)$, extending \cite{gahrooei2019process} to include the intercept function and retaining only the roughness penalty, as variable selection is a Phase I concern.
Following \cite{gahrooei2019process}, let $\boldsymbol{\phi}(t) = \left(\phi_1(t),\dots,\phi_{k_y}(t)\right)^T \in \mathbb{R}^{k_y}$ and $\boldsymbol{\theta}_j(s) = \left(\theta_{j1}(s),\dots, \theta_{jk_{x_j}}(s) \right)^T \in \mathbb{R}^{k_{x_j}}$, $j = 1,\dots,p$, be vectors of basis functions evaluated at points $t$ and $s$, and corresponding to the functional response $Y(t)$ and the functional covariates $\boldsymbol{X}(s)$, respectively. Using these basis functions, the infinite-dimensional estimation problem can be reduced to a finite-dimensional problem by expanding the intercept function and the regression coefficient functions as follows
\begin{equation*}
    \beta_{0i}(t) = \sum_{k = 1}^{k_y}b_{0ik}\phi_k(t),
    \quad \beta_{ji}(s,t) = \sum_{k = 1}^{k_y}\sum_{q = 1}^{k_{x_j}}b_{jikq}\theta_{jq}(s)\phi_k(t), \quad s \in \mathcal{S}, \quad t \in \mathcal{T}, \quad i = 1,2,\dots.
\end{equation*}
Then, Equation \eqref{flm} can be written as
\begin{equation}
\label{flm_exp}
    Y_i(t) =  \boldsymbol{b}_{0i}^T\boldsymbol{\phi}(t) + \sum_{j = 1}^p\int_\mathcal{S}X_{ji}(s)\boldsymbol{\theta}_j^T(s)\boldsymbol{B}_{ji}\boldsymbol{\phi}(t)ds + \varepsilon_i(t), \quad t \in \mathcal{T},\quad i = 1,2,\dots,
\end{equation}
where $\boldsymbol{b}_{0i} = \left(b_{0i1},\dots,b_{0ik_y} \right)^T$ is the vector of coefficients corresponding to $\beta_{0i}(t)$, and $\boldsymbol{B}_{ji} = \left\{b_{jikq}\right\}_{k = 1,\dots k_y, q = 1,\dots,k_{x_j}}$, $j = 1,\dots,p$, is the matrix of coefficients of dimension $k_{x_j} \times k_y$ corresponding to the regression coefficient functions $\beta_{ji}(s,t)$, $j = 1,\dots,p$.
Moreover, let us define $\boldsymbol{Z}_{ji} \coloneq \left(\int_\mathcal{S}X_{ji}(s) \boldsymbol{\theta}^T_j(s)ds\right)^T \in \mathbb{R}^{k_{x_j}}$, $\boldsymbol{Y}_i \coloneq \left(Y_i(t_1),\dots,Y_i(t_n)\right)^T \in \mathbb{R}^n$, $\boldsymbol{\Phi} = \left(\boldsymbol{\phi}(t_1),\dots,\boldsymbol{\phi}(t_n)\right) \in \mathbb{R}^{k_y \times n}$, and $\boldsymbol{E}_i = \left(\varepsilon_i(t_1),\dots,\varepsilon_i(t_n)\right)^T \in \mathbb{R}^n$.
Then, Equation \eqref{flm_exp} can be written as
\begin{equation*}
    \boldsymbol{Y}_i = \boldsymbol{\Phi}^T\boldsymbol{b}_{0i} + \sum_{j= 1}^p \left(\boldsymbol{Z}_{ji}^T\boldsymbol{B}_{ji}\boldsymbol{\Phi}\right)^T + \boldsymbol{E}_i, \quad i = 1,2,\dots,
\end{equation*}
which can be further simplified by introducing the quantity $\boldsymbol{D}_{ji} = \boldsymbol{Z}_{ji} \otimes \boldsymbol{\Phi} \in \mathbb{R}^{k_{x_j}k_y \times n}$ and considering $\boldsymbol{b}_{ji} = \mathrm{vec}\left(\boldsymbol{B}_{ji}^T\right)$ as follows
\begin{align}
\label{flm_red_1}
    \boldsymbol{Y}_i & = \boldsymbol{\Phi}^T\boldsymbol{b}_{0i} + \sum_{j= 1}^p \left(\boldsymbol{Z}_{ji} \otimes \boldsymbol{\Phi}\right)^T\boldsymbol{b}_{ji} + \boldsymbol{E}_i \nonumber \\
    & =  \boldsymbol{\Phi}^T\boldsymbol{b}_{0i} + \sum_{j= 1}^p \boldsymbol{D}_{ji}^T\boldsymbol{b}_{ji} + \boldsymbol{E}_i, \quad i = 1,2,\dots.
\end{align}
The $\mathrm{vec}(\boldsymbol{X})$ operator stacks the columns of a matrix $\boldsymbol{X}$ into a vector.
Equation \eqref{flm_red_1} can then be written compactly as
\begin{equation}
    \label{flm_red_2}
        \boldsymbol{Y}_i =\boldsymbol{D}_i\boldsymbol{b}_i + \boldsymbol{E}_i,\quad i = 1,2,\dots,
\end{equation}
with $\boldsymbol{D}_i = \left(\boldsymbol{\Phi}^T,\boldsymbol{D}_{1i}^T,\dots,\boldsymbol{D}_{pi}^T\right) \in \mathbb{R}^{n \times k_y\left(1+\sum_{j = 1}^pk_{x_j}\right)}$, and $\boldsymbol{b}_i = \left(\boldsymbol{b}_{0i}^T,\boldsymbol{b}_{1i}^T,\dots,\boldsymbol{b}_{pi}^T\right)^T$.
Thus, we can estimate the coefficient vector $\boldsymbol{b}_i$ as follows.
Let 
\begin{equation*}
\ell^*\left(\boldsymbol{b}_i;\boldsymbol{D}_i,\boldsymbol{Y}_i\right)=\frac{1}{2}\left\|\boldsymbol{Y}_i- \boldsymbol{D}_i\boldsymbol{b}_i\right\|_2^2 = \frac{1}{2}\boldsymbol{b}_i^T\boldsymbol{D}_i^T\boldsymbol{D}_i\boldsymbol{b}_i - \boldsymbol{Y}_i^T\boldsymbol{D}_i\boldsymbol{b}_i + \frac{1}{2}\boldsymbol{Y}_i^T\boldsymbol{Y}_i, \quad i = 1,2,\dots
\end{equation*}
be the quadratic loss function corresponding to the functional observation $\left(Y_i,\boldsymbol{X}_i \right)$ and note that it is generally infeasible to minimize for a single observation, because the number of parameters $k_y\left(1 + \sum_{j=1}^p k_{x_j}\right)$ typically exceeds $n$, yielding an underdetermined system.
One natural solution could be to incorporate past observations. 
To increase the sensitivity in detecting an OC state of the process, we therefore assign unequal weights to past observations and use the following exponentially weighted moving loss \citep{luxenberg2024exponentially}
\begin{equation}
    \label{loss}
       \ell(\boldsymbol{b}_i) = \omega_i \sum_{\tau = 1}^i\eta^{i-\tau}\ell^*(\boldsymbol{b}_i;\boldsymbol{D}_\tau,\boldsymbol{Y}_\tau) + r(\boldsymbol{b}_i),\quad i = 1,2,\dots,
\end{equation}
where $\eta \in (0,1)$ is a forgetting factor, and
\begin{equation*}
    \omega_i = \left(\sum_{\tau = 1}^i \eta^{i-\tau}\right)^{-1} = \frac{1 - \eta}{ 1 - \eta^i}
\end{equation*}
is the normalization constant.
The loss function in Equation \eqref{loss} gives more weight to recent observations through the exponential factor $\eta^{i-\tau}$, so that the weight assigned to observation $\tau$ at sampling stage $i$ decreases exponentially as 
$i$ increases, and the most recent observation receives the highest weight. The forgetting factor $\eta$ controls the decay rate, with values close to $1$ yielding nearly equal weights across past observations. In contrast, smaller values concentrate the weight on the most recent ones.
The term $r(\boldsymbol{b}_i) = \frac{\lambda}{2}\sum_{j=0}^p \boldsymbol{b}_{ji}^T \boldsymbol{R}_j \boldsymbol{b}_{ji}$ is a roughness penalty, where $\lambda > 0$ is a tuning parameter and $\boldsymbol{R}_j$ is a roughness penalty matrix that depends on $j$. 
In particular, for $j = 1,\dots,p$, the roughness penalty matrix is $\boldsymbol{R}_j = \boldsymbol{R}_{j,s} + \boldsymbol{R}_{j,t}$, where $\boldsymbol{R}_{j,s} = \left[\int_\mathcal{S}L_s\boldsymbol{\theta}_j(s)
L_s\boldsymbol{\theta}_j^T(s)ds\right] \otimes \left[\int_\mathcal{T}
\boldsymbol{\phi}(t)\boldsymbol{\phi}^T(t)dt\right]$ and $\boldsymbol{R}_{j,t} = 
\left[\int_\mathcal{S}\boldsymbol{\theta}_j(s)\boldsymbol{\theta}_j^T(s)ds
\right] \otimes \left[\int_\mathcal{T}L_t\boldsymbol{\phi}(t)
L_t\boldsymbol{\phi}^T(t)dt\right]$, with $L_s$ and $L_t$ being differentiating operators with respect to $s$ and $t$ \citep{gahrooei2019process}.
The penalties $\boldsymbol{R}_{j,s}$ and $\boldsymbol{R}_{j,t}$ penalize the roughness of $\beta_{ji}(s,t)$ in the $s$ and $t$ directions, respectively, encouraging smooth estimates of the regression coefficient functions.
For $j = 0$, the intercept function $\beta_{0i}(t)$ depends only on $t$, and the penalty reduces to $\boldsymbol{R}_0 = \int_\mathcal{T}L_t\boldsymbol{\phi}(t)
L_t\boldsymbol{\phi}^T(t)dt$.

For a quadratic loss, as in our case, \cite{luxenberg2024exponentially} showed that the parameter to be estimated can be computed recursively using a simple recursion similar to the exponentially weighted moving average (EWMA) \citep{lucas1990exponentially}, which allows us to carry out computations of constant complexity, storing only a fixed-size state.
To be specific, let $P(\boldsymbol{D}_i) = \boldsymbol{D}_i^T\boldsymbol{D}_i$, $p(\boldsymbol{D}_i,\boldsymbol{Y}_i) = -\boldsymbol{D}_i^T\boldsymbol{Y}_i$ and $\pi(\boldsymbol{Y}_i) = \frac{1}{2}\boldsymbol{Y}_i^T\boldsymbol{Y}_i$.
Equation \eqref{loss} can be written as \citep{luxenberg2024exponentially} 
\begin{equation}
    \label{loss2}
       \ell(\boldsymbol{b}_i) = \frac{1}{2}\boldsymbol{b}_i^T\boldsymbol{P}_i\boldsymbol{b}_i + \boldsymbol{p}_i^T\boldsymbol{b}_i  + \pi_i + r(\boldsymbol{b}_i),\quad i = 1,2,\dots,
\end{equation}
with
\begin{equation*}
\boldsymbol{P}_i=\omega_i \sum_{\tau=1}^i \eta^{i-\tau} P\left(\boldsymbol{D}_\tau\right), \quad \boldsymbol{p}_i=\omega_i \sum_{\tau=1}^i \eta^{i-\tau} p\left(\boldsymbol{D}_\tau,\boldsymbol{Y}_\tau\right), \quad \pi_i=\omega_i \sum_{\tau=1}^i \eta^{i-\tau} \pi\left(\boldsymbol{Y}_\tau\right),
\end{equation*}
which can be stored using the recursions
\begin{align*}
\boldsymbol{P}_{i} & =\frac{\omega_i}{\omega_{i-1}} \eta \boldsymbol{P}_{i-1}+\omega_i P\left(\boldsymbol{D}_{i}\right), \\
\boldsymbol{p}_{i} & =\frac{\omega_i}{\omega_{i-1}} \eta \boldsymbol{p}_{i-1}+\omega_i p\left(\boldsymbol{D}_{i},\boldsymbol{Y}_i\right), \\
\pi_{i} & =\frac{\omega_i}{\omega_{i-1}} \eta \pi_{i-1}+\omega_i \pi\left(\boldsymbol{Y}_{i}\right),
\end{align*}
with $\boldsymbol{P}_0$, $\boldsymbol{p}_0$ and $\pi_0$ computed using IC observations in the Phase I sample.
Then, $\boldsymbol{b}_i$ is computed by minimizing Equation \eqref{loss2}, whose solution is given by
\begin{equation*}
    \widehat{\boldsymbol{b}}_i = -\left(\boldsymbol{P}_i + \lambda \boldsymbol{R} \right)^{-1}\boldsymbol{p}_i,
\end{equation*}
where $\boldsymbol{R}$ is a block diagonal matrix with blocks $\boldsymbol{R}_0, \dots, \boldsymbol{R}_p$.
Note that $\boldsymbol{P}_i + \lambda\boldsymbol{R}$ is invertible for every $i$, since the Phase~I sample is large enough to make $\boldsymbol{P}_0$ positive definite. The recursion has strictly positive coefficients and adds positive semidefinite terms, so it propagates positive definiteness to all subsequent sampling stages.
Given $\widehat{\boldsymbol{b}}_i = \left(\widehat{\boldsymbol{b}}_{0i}^T, 
\widehat{\boldsymbol{b}}_{1i}^T, \dots, \widehat{\boldsymbol{b}}_{pi}^T\right)^T$, the estimated intercept function and regression coefficient functions at sampling stage $i$ are recovered as \begin{equation*}
    \widehat{\beta}_{0i}(t) = \widehat{\boldsymbol{b}}_{0i}^T\boldsymbol{\phi}(t), 
    \qquad
    \widehat{\beta}_{ji}(s,t) = \boldsymbol{\theta}_j^T(s)\widehat{\boldsymbol{B}}_{ji}
    \boldsymbol{\phi}(t), \quad j = 1,\dots,p,
\end{equation*}
where $\widehat{\boldsymbol{B}}_{ji}$ is obtained by reshaping $\widehat{\boldsymbol{b}}_{ji} = \mathrm{vec}\left(\widehat{\boldsymbol{B}}_{ji}^T\right)$, 
yielding the estimated stacked coefficient vector  $\widehat{\boldsymbol{\beta}}_i = \left(\widehat{\beta}_{0i}, \widehat{\beta}_{1i}, \dots, \widehat{\beta}_{pi}\right)^T$, which is used to test the hypothesis in Equation \eqref{hyp} as described in Section \ref{monitoring}.

\subsection{Monitoring Procedure}
\label{monitoring}
At each sampling stage $i$, the estimated stacked coefficient vector $\widehat{\boldsymbol{\beta}}_i = \left(\widehat{\beta}_{0i}, \widehat{\beta}_{1i}, \dots, \widehat{\beta}_{pi}\right)^T$ obtained in Section \ref{seqest} lies in the finite-dimensional space defined by the chosen basis functions and is determined by the coefficient vector $\widehat{\boldsymbol{b}}_i$. Nevertheless, its dimension $k_y\left(1 + \sum_{j=1}^p k_{x_j}\right)$ may be large, and its components, defined over different dimensional domains, are usually correlated and exhibit heterogeneous variability. Note also that the true stacked coefficient vector $\boldsymbol{\beta}_0$ is fixed. In contrast, its estimator $\widehat{\boldsymbol{\beta}}_i$ is random, with variability induced by the randomness in the observations $(Y_i, \boldsymbol{X}_i)$ and by the finite-sample nature of the sequential estimator. Let $\overline{\boldsymbol{\beta}} = \left(\overline{\beta}_0, \overline{\beta}_1, \dots, \overline{\beta}_p\right)^T$ denote the IC mean of $\widehat{\boldsymbol{\beta}}_i$. Although the hypothesis in Equation \eqref{hyp} concerns the deviation $\boldsymbol{\beta}_i - \boldsymbol{\beta}_0$, centering on the IC mean $\overline{\boldsymbol{\beta}}$ removes any fixed offset, so that any departure of $\boldsymbol{\beta}_i$ from $\boldsymbol{\beta}_0$ manifests as a departure of $\widehat{\boldsymbol{\beta}}_i$ from $\overline{\boldsymbol{\beta}}$, which the monitoring statistics defined below are designed to detect.
To balance the contribution of each component to the analysis, following \citet{happ2018multivariate}, the centered components are rescaled as
\begin{equation}
\label{rescaled}
\widetilde{\beta}_{ji} = w_j^{1/2}\left(\widehat{\beta}_{ji} - \overline{\beta}_j\right), \qquad w_j = \left(\int_{\mathcal{D}_j} \mathrm{Var}\big(\widehat{\beta}_{ji}(\boldsymbol{u})\big)\,d\boldsymbol{u}\right)^{-1}, \quad j = 0, 1, \dots, p,
\end{equation}
where $\mathcal{D}_0 = \mathcal{T}$ and $\mathcal{D}_j = \mathcal{S}\times \mathcal{T}$ for $j \geq 1$. With this choice of weights, the integrated variance of each rescaled component equals one, so all components contribute equally to the analysis. From the multivariate Karhunen--Lo\`eve theorem \citep{happ2018multivariate}, the rescaled stacked coefficient vector $\widetilde{\boldsymbol{\beta}}_i = \left(\widetilde{\beta}_{0i}, \widetilde{\beta}_{1i}, \dots, \widetilde{\beta}_{pi}\right)^T$ admits the expansion
\begin{equation}
\label{kl}
    \widetilde{\boldsymbol{\beta}}_i = \sum_{l=1}^\infty \xi_{il}\boldsymbol{\psi}_l,
\end{equation}
where $\boldsymbol{\psi}_l = \left(\psi_{0l}, \psi_{1l}, \dots, \psi_{pl}\right)^T$ are the multivariate principal components, that is, the eigenfunctions of the covariance of $\widetilde{\boldsymbol{\beta}}_i$ associated with the eigenvalues $\zeta_1 \geq \zeta_2 \geq \dots \geq 0$, and $\xi_{il} = \langle\widetilde{\boldsymbol{\beta}}_i, \boldsymbol{\psi}_l \rangle_{\mathbb{H}_\beta}$ are the principal component scores, such that $\E(\xi_{il}) = 0$ and $\E(\xi_{il}\xi_{im}) = \zeta_l\delta_{lm}$, where $\delta_{lm}$ is the Kronecker delta. As the eigenvalues approach zero, the leading principal components are assumed to capture the most important features of the variability in $\widetilde{\boldsymbol{\beta}}_i$. Therefore, by retaining the $L$ leading principal components to explain at least a prespecified percentage of the total variability \citep{paynabar2016change, ren2019phase, centofanti2021bfunctional, capezza2026functional}, the rescaled stacked coefficient vector in Equation \eqref{kl} is approximated through the truncated decomposition $\widetilde{\boldsymbol{\beta}}_i^L = \sum_{l=1}^L \xi_{il}\boldsymbol{\psi}_l$.

The functional extension of the classical Hotelling's statistic, based on the rescaled stacked coefficient vector $\widetilde{\boldsymbol{\beta}}_i$, can then be defined as
\begin{equation}
\label{t2}
    T^2_i = \sum_{j=0}^p \sum_{k=0}^p \int_{\mathcal{D}_j} \int_{\mathcal{D}_k} 
    \widetilde{\beta}_{ji}(\boldsymbol{u}) \, K^*_{jk}(\boldsymbol{u}, 
    \boldsymbol{v}) \, \widetilde{\beta}_{ki}(\boldsymbol{v}) \, d\boldsymbol{u} 
    \, d\boldsymbol{v},
\end{equation}
with
\begin{equation}
\label{kernel}
    K^*_{jk}(\boldsymbol{u}, \boldsymbol{v}) = \sum_{l=1}^L \frac{1}{\zeta_l} 
    \psi_{jl}(\boldsymbol{u}) \psi_{kl}(\boldsymbol{v}), \qquad j,k = 0,1,\dots,p,
\end{equation}
where $\psi_{jl}$ denotes the $j$-th component of the $l$-th principal component $\boldsymbol{\psi}_l$. The $T^2_i$ statistic in Equation~\eqref{t2} is the functional extension of the likelihood-ratio test statistic for the mean of multivariate normal data \citep{johnson2002applied}, and, exploiting the orthonormality of the principal components, it can be equivalently written in terms of the principal component scores as $T^2_i = \sum_{l=1}^L \xi_{il}^2/\zeta_l$.
Analogously, changes that may occur in the subspace orthogonal to the one spanned by the first $L$ principal components are monitored through the $SPE$ statistic at each sampling stage $i = 1,2,\dots$
\begin{equation}
\label{spe}
    SPE_i = \left\|\widetilde{\boldsymbol{\beta}}_i - 
    \widetilde{\boldsymbol{\beta}}_i^L\right\|_{\mathbb{H}_\beta}^2 = 
    \sum_{j=0}^p \int_{\mathcal{D}_j} \left(\widetilde{\beta}_{ji}(\boldsymbol{u}) 
    - \widetilde{\beta}_{ji}^L(\boldsymbol{u})\right)^2 d\boldsymbol{u},
\end{equation}
where $\widetilde{\beta}_{ji}^L$ is the $j$-th component of the truncated reconstruction $\widetilde{\boldsymbol{\beta}}_i^L$ and the difference quantifies the reconstruction error of the truncated representation.

An OC signal is issued if $T^2_i > h_{T^2}$ or $SPE_i > h_{SPE}$, where $h_{T^2}$ and $h_{SPE}$ are the corresponding control limits calibrated to target a prespecified $ARL_0$. Section \ref{implementation} provides more details on estimating the control limits.

\subsection{Post-signal Diagnostics}
\label{post-signal}
Once an OC signal is issued, it is often useful to identify which components of the stacked coefficient vector drive the detected change. To this end, we propose a diagnostic procedure based on the decomposition of the monitoring statistics into the contributions of the 
individual components of $\widehat{\boldsymbol{\beta}}_i$.

The additive structure of Hotelling's statistic $T^2_i$ in Equation \eqref{t2} allows the following decomposition 
\begin{equation*}
    T^2_i = \sum_{l=1}^L \frac{\xi_{il}^2}{\zeta_l} =\sum_{j=0}^p \sum_{l=1}^L 
    \frac{\xi_{il}}{\zeta_l} \langle\widetilde{\beta}_{ji}, 
    \psi_{jl}\rangle_{\mathcal{D}_j},
\end{equation*}
and accordingly, the definition of the following contributions
\begin{equation*}
    c^{T^2}_{ji} = \sum_{l=1}^L 
    \frac{\xi_{il}}{\zeta_l} \langle\widetilde{\beta}_{ji}, 
    \psi_{jl}\rangle_{\mathcal{D}_j}, \quad j = 0,\dots,p.
\end{equation*}
Analogously, the contributions to the $SPE_i$ statistic in Equation \eqref{spe} can be defined as
\begin{equation*}
    c^{SPE}_{ji} =  \int_{\mathcal{D}_j} 
    \left(\widetilde{\beta}_{ji}(\boldsymbol{u}) - 
    \widetilde{\beta}_{ji}^L(\boldsymbol{u})\right)^2 
    d\boldsymbol{u}, \quad j = 0,\dots,p.
\end{equation*}
Following the idea of \citet{westerhuis2000generalized}, also adopted by \cite{centofanti2025adaptive}, we introduce control limits for individual contributions to help identify anomalous components.  Specifically, the $j$-th component is flagged as responsible for the OC condition when its contribution exceeds the corresponding threshold, namely, $c^{T^2}_{ji} > h^{T^2}_j$ or $c^{SPE}_{ji} > h^{SPE}_j$. Thresholds $h^{T^2}_j$ and $h^{SPE}_j$ are obtained as the $\left(1 - \alpha_j \right)$ quantiles of the IC marginal distributions of $c^{T^2}_{ji}$ and $c^{SPE}_{ji}$, with $\alpha_j$ denoting the overall Type I error probability corresponding to $\widehat{\beta}_{ji}$, $j = 0,\dots,p$, $i = 1,2,\dots$.
Of course, if the OC signal is issued by $T^2_i$ (resp. $SPE_i$), we examine only the contributions $c^{T^2}_{ji}$ (resp. $c^{SPE}_{ji}$).

\subsection{Implementation Details}
\label{implementation}
The functional response and covariates are usually observed on discrete grids, and a Phase I sample of $N$ IC observations $\{(Y_i, \boldsymbol{X}_i)\}_{i=1}^N$ to characterize the IC state of the process is assumed to be available. The Phase I sample is partitioned into a training set and a tuning set. The training set is used to estimate the IC stacked coefficient vector $\widehat{\boldsymbol{\beta}}_0$, to select the roughness parameter $\lambda$, and to estimate the MFPCA model and the IC mean $\overline{\boldsymbol{\beta}}$; whereas the tuning set is used to estimate the control limits and reduce overfitting in their estimation \citep{kruger2012statistical}. Following \cite{gahrooei2019process}, the IC stacked coefficient vector $\widehat{\boldsymbol{\beta}}_0$ is estimated from the centered training set and is used both to initialize the sequential estimation and as the reference against which $\lambda$ is selected.

The basis functions $\boldsymbol{\phi}(t)$ and $\boldsymbol{\theta}_j(s)$, $j = 1,\dots,p$, are chosen as B-spline bases, which are a common choice for non-periodic functional data due to their flexibility and favorable computational properties \citep{gahrooei2019process, ramsay}. The numbers of basis functions $k_y$ and $k_{x_j}$, $j = 1,\dots,p$, should be large enough to capture the local behavior of the functional response and covariates \citep{cardot2003spline}. At the same time, the roughness penalty controls the smoothness of the estimates. In our implementation, we use cubic B-splines and compute the roughness penalties $\boldsymbol{R}_j$, $j = 0,\dots,p$ by penalizing the integrated squared second derivative. Supplementary Materials~\ref{supp_nbasis} provides a sensitivity analysis of the number of basis functions.

The forgetting factor $\eta \in (0,1)$ governs the trade-off between the sensitivity of the sequential estimates to recent observations and the stability of the estimation, and $1-\eta$ can be envisaged as playing a role that is similar to that of the weighting parameter in the EWMA control chart \citep{lucas1990exponentially}. Small values of $\eta$ concentrate the weight on the most recent observations, enhancing responsiveness to abrupt changes, whereas values close to 1 yield smoother estimates and therefore greater sensitivity to small, persistent shifts. In the absence of prior information on the magnitude of the change, $\eta$ is fixed to a value close to one, such as $0.95$, corresponding to an EWMA smoothing parameter $1-\eta = 0.05$, since the proposed control chart is primarily intended to detect small and persistent departures from the IC state. Supplementary Materials~\ref{supp_eta} provides a sensitivity analysis of the effect of $\eta$.

The roughness parameter $\lambda$ is selected so that, under IC conditions, the sequentially estimated stacked coefficient vectors $\widehat{\boldsymbol{\beta}}_i, i = 1,2,\dots,$ remain as close as possible to the Phase I estimate $\widehat{\boldsymbol{\beta}}_0$. Specifically, several IC sequences of fixed length are generated by randomly sampling from the training set, the corresponding sequential estimates $\widehat{\boldsymbol{\beta}}_i$ are computed, and $\lambda$ is chosen as the value minimizing the average discrepancy $\|\widehat{\boldsymbol{\beta}}_i - \widehat{\boldsymbol{\beta}}_0\|_{\mathbb{H}_\beta}^2$ across the sequences and the observations within them. The rationale is that, under $H_0$, the sequential estimates should reproduce the IC relationship encoded in $\widehat{\boldsymbol{\beta}}_0$.

The MFPCA model is estimated from a sample of sequential estimates of the stacked coefficient vector $\widehat{\boldsymbol{\beta}}_i$ obtained by generating IC sequences of fixed length from the training set, from which the IC mean $\overline{\boldsymbol{\beta}}$ and the weights $w_j$ of Equation \eqref{rescaled} are computed. Following the basis function expansion approach of \cite{ramsay}, each component $\psi_{jl}$ of the principal components $\boldsymbol{\psi}_l$ is represented through the same basis functions used to obtain $\widehat{\boldsymbol{\beta}}_i$, that is, $\psi_{0l}(t) = \boldsymbol{c}_{0l}^T\boldsymbol{\phi}(t)$ and $\psi_{jl}(s,t) = \boldsymbol{\theta}_j^T(s)\boldsymbol{C}_{jl}\boldsymbol{\phi}(t)$, $j = 1,\dots,p$, $l = 1, 2, \dots$, where $\boldsymbol{c}_{0l}$ and $\boldsymbol{C}_{jl}$ are the corresponding eigenfunction coefficients. Under these assumptions, MFPCA \citep{happ2018multivariate, ramsay} consists of performing a standard multivariate principal component analysis of the rescaled coefficient vectors $\boldsymbol{W}^{1/2}\widetilde{\boldsymbol{b}}_i$, where $\widetilde{\boldsymbol{b}}_i = \left(\widetilde{\boldsymbol{b}}_{0i}^T,\widetilde{\boldsymbol{b}}_{1i}^T, \dots, \widetilde{\boldsymbol{b}}_{pi}^T\right)^T$, with $\widetilde{\boldsymbol{b}}_{ji} = w_j^{1/2}\left(\widehat{\boldsymbol{b}}_{ji}-\bar{\boldsymbol{b}}_{j}\right)$, $j = 0,\dots,p$, is the vector of coefficients corresponding to the standardized vector $\widetilde{\boldsymbol{\beta}}_i$ defined in Equation \eqref{rescaled}, and $\boldsymbol{W}$ is the block-diagonal matrix with diagonal blocks $\boldsymbol{W}_j$, $j = 0,\dots,p$, whose entries are the inner products of the corresponding basis functions. 
Specifically, the estimated eigenvalues $\zeta_l$ are the eigenvalues of the sample covariance matrix $\boldsymbol{S}$ obtained from $\boldsymbol{W}^{1/2}\widetilde{\boldsymbol{b}}_i$, while the eigenfunction coefficients $\boldsymbol{c}_l$ are obtained as $\boldsymbol{c}_l = \boldsymbol{W}^{-1/2}\boldsymbol{u}_l$, where $\boldsymbol{u}_l$ is the $l$-th eigenvector of $\boldsymbol{S}$. We choose the number of retained components $L$ so that the explained variability reaches a prescribed fraction, as discussed in Section \ref{monitoring}. Supplementary Materials~\ref{supp_fve} provides the sensitivity analysis.

The control limits $h_{T^2}$ and $h_{SPE}$, as well as the contribution limits $h^{T^2}_j$ and $h^{SPE}_j$, $j = 0,\dots,p$, are estimated through a bootstrap procedure on the tuning set \citep{capezza2026functional}. Specifically, we generate $n_{\mathrm{seq}}$ sequences of length $l_{\mathrm{seq}}$ by randomly sampling without replacement from the tuning set. 
For all sequences, we compute $T^2$ and $SPE$ monitoring statistics. The run length is recorded as the first observation that exceeds the current tentative values of $h_{T^2}$ and/or $h_{SPE}$, initialized as the maximum of the corresponding  
$n_{\mathrm{seq}} \times l_{\mathrm{seq}}$ values of the statistic.
We iteratively adjust the control limits using the bisection algorithm for multi-chart designs \citep{zago2025improved} until the estimated $ARL_0$ reaches a prespecified value. To avoid discarding sequences that do not produce a signal, we adopt the approach of \citet{lim2024efficient}, retaining non-signaling sequences as Type I censored observations when estimating $ARL_0$. We obtain the contribution limits $h^{T^2}_j$ and $h^{SPE}_j$ as the $(1-\alpha_j)$ quantiles of the empirical IC marginal distributions of the corresponding contributions computed on the tuning set.

\section{Simulation Study}
\label{sec_sim}
\begin{figure}
    \centering
    \includegraphics[width=1\linewidth]{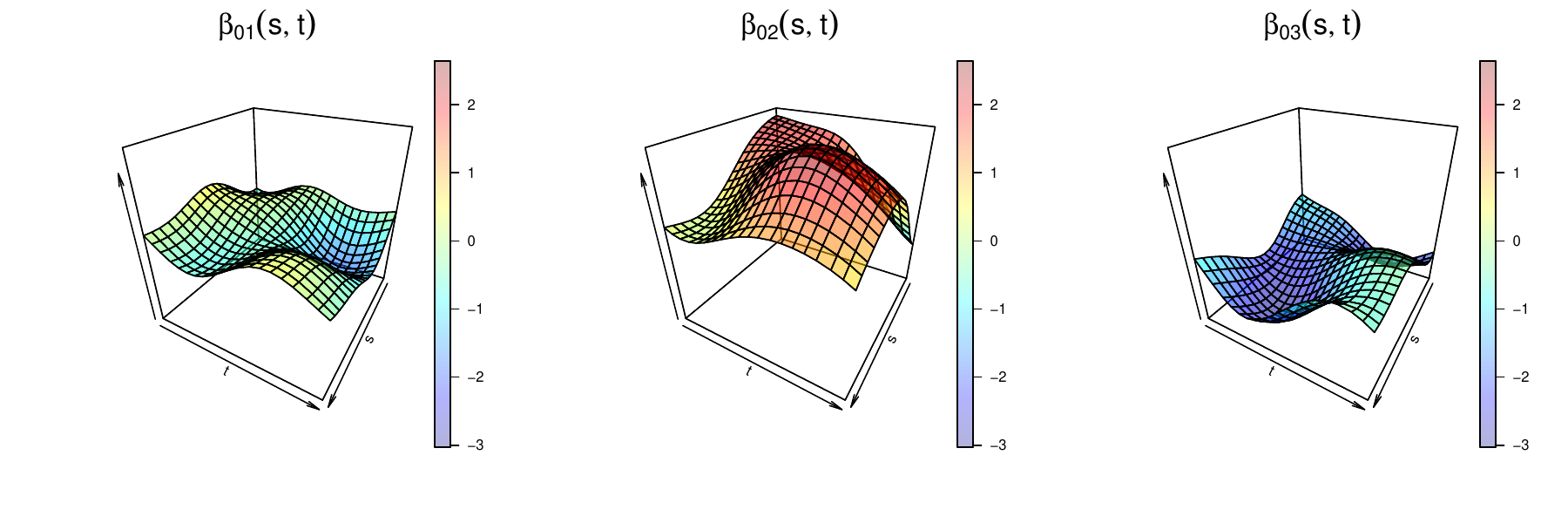}
    \caption{IC regression coefficient functions $\beta_{0j}(s,t)$ corresponding to the first three functional covariates used in the simulation study.}
    \label{fig:betas}
\end{figure}

To evaluate the performance of the proposed RCF control chart in detecting a departure from the IC state of the process, the IC functional response has been generated according to the following FLM
\begin{equation}
    \label{flm_ic_sim}
    Y(t) = \sum_{j = 1}^p\int_{[0,1]}\beta_{0j}(s,t)X_j(s)ds + \varepsilon(t), \quad t \in [0,1],
\end{equation}
where, without loss of generality, the regression coefficient functions $\beta_{0j}$, $j = 1,\dots,p$, are generated as smooth bivariate surfaces over $[0,1] \times [0,1]$, each drawn from a two-dimensional Gaussian process with a separable squared-exponential covariance kernel. A detailed description of the generation of the regression coefficient functions $\beta_{0j}$, $j = 1,\dots,p$, the functional covariates $X_j$, and the error term $\varepsilon$ is provided in Supplementary Materials~\ref{supp_datagen}. As an illustration, Figure \ref{fig:betas} depicts the coefficient functions $\beta_{0j}$, $j = 1,\dots,3$, corresponding to the first three functional covariates.

We consider three scenarios, each corresponding to an increasing number of functional covariates ($p = 1,3,5$). In each scenario, we explore three different shift types in the regression coefficient function $\beta_{01}(s,t)$ and, separately, in the intercept function $\beta_{00}(t)$. 
In particular, Shift A is characterized by a decrease in the regression coefficient function (along the $s$ direction) or in the mean function in the central part of the domain. Shift B is characterized by a sinusoidal behavior. Shift C corresponds to a quadratic decrease in the regression coefficient function (along the $s$ direction) or in the mean function, respectively. For each shift, we consider 11 increasing severity levels $SL$, where $SL=0$ corresponds to the IC state.
\begin{figure}
    \centering
    \includegraphics[width=1\linewidth]{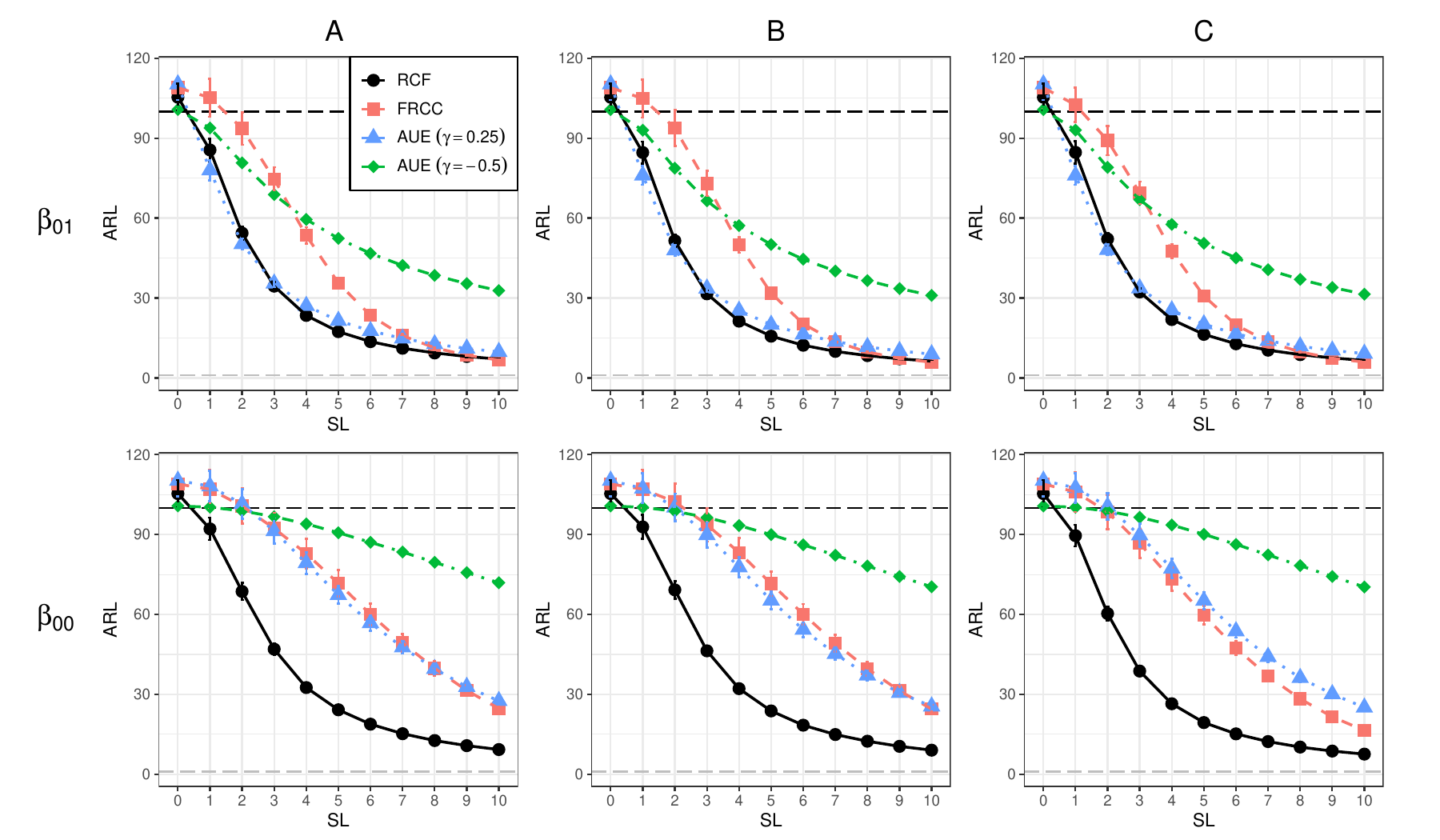}
    \caption{Mean Phase II $ARL$ over the Monte Carlo simulation runs and the corresponding $95\%$ confidence interval achieved by RCF, FRCC and AUE  as a function of the severity level SL, for $p = 1$ functional covariate.}
    \label{fig:p_1}
\end{figure}
We compare the proposed method with the method of \cite{aue2014dependent}, referred to as AUE, and the functional regression control chart of \cite{centofanti2021bfunctional}, referred to as FRCC. As noted in the Introduction, AUE applies only to $p = 1$ and is therefore included only in the corresponding scenario. It is worth noting that AUE is not natively designed to attain a prespecified $ARL_0$, and requires choosing the curved threshold hyperparameter $\gamma$, which governs the sensitivity of detection across different change-point locations. 
In contrast, the proposed method relies on a constant control limit, calibrated to achieve a prespecified $ARL_0$, and, once the steady state is reached, its detection performance does not depend on the change-point location, as the exponentially weighted moving loss in
Equation \eqref{loss} assigns weights that depend only on the lag $i - \tau$ and not on the sampling stage itself.
To enable a fair comparison, we compute the control limit using the same bootstrap procedure defined in Section \ref{implementation} to achieve the same $ARL_0$ across methods. 
We study cases with $\gamma =- 0.5$ and $\gamma = 0.25$. 
In particular, $\gamma = -0.5$ was used by \cite{aue2014dependent} in their simulation study, while $\gamma = 0.25$ yields more balanced detection across change-point locations when the change time is unknown.
For each combination of scenarios, shift types, and $SL$, we run $100$ Monte Carlo simulations. As in Section \ref{implementation}, each run independently generates a training set and a tuning set with 500 and 1000 IC functional observations, respectively, to mimic a Phase I sample. Both the functional response and the functional covariates are standardized following the approach of \cite{chiou2014multivariate}, as in \cite{centofanti2021bfunctional}. The training sample is used to estimate the IC regression coefficient functions using the method proposed in \cite{gahrooei2019process} with no variable-selection penalty, and the MFPCA parameters are estimated by generating $n_{\mathrm{seq}} = 200$ sequences of length $l_{\mathrm{seq}} = 500$. We use the same $n_{\mathrm{seq}}$ and $l_{\mathrm{seq}}$ to generate sequences from the tuning set, from which we compute the control limits and contribution limits. We set the number of B-spline bases to 15 for both the functional response and the functional covariates, and set the forgetting factor to $\eta = 0.95$. We set the fraction of variance explained (FVE) to 0.95 for all methods.
In all scenarios, we set the change point at the beginning of the monitoring phase. For the proposed method, a sequence of 200 IC observations precedes the change point, so that steady-state
conditions are attained before monitoring begins. AUE and FRCC need no warm-up sequence, as they do not involve a recursive state and are therefore unaffected by initialization.
\begin{figure}
    \centering
    \includegraphics[width=1\linewidth]{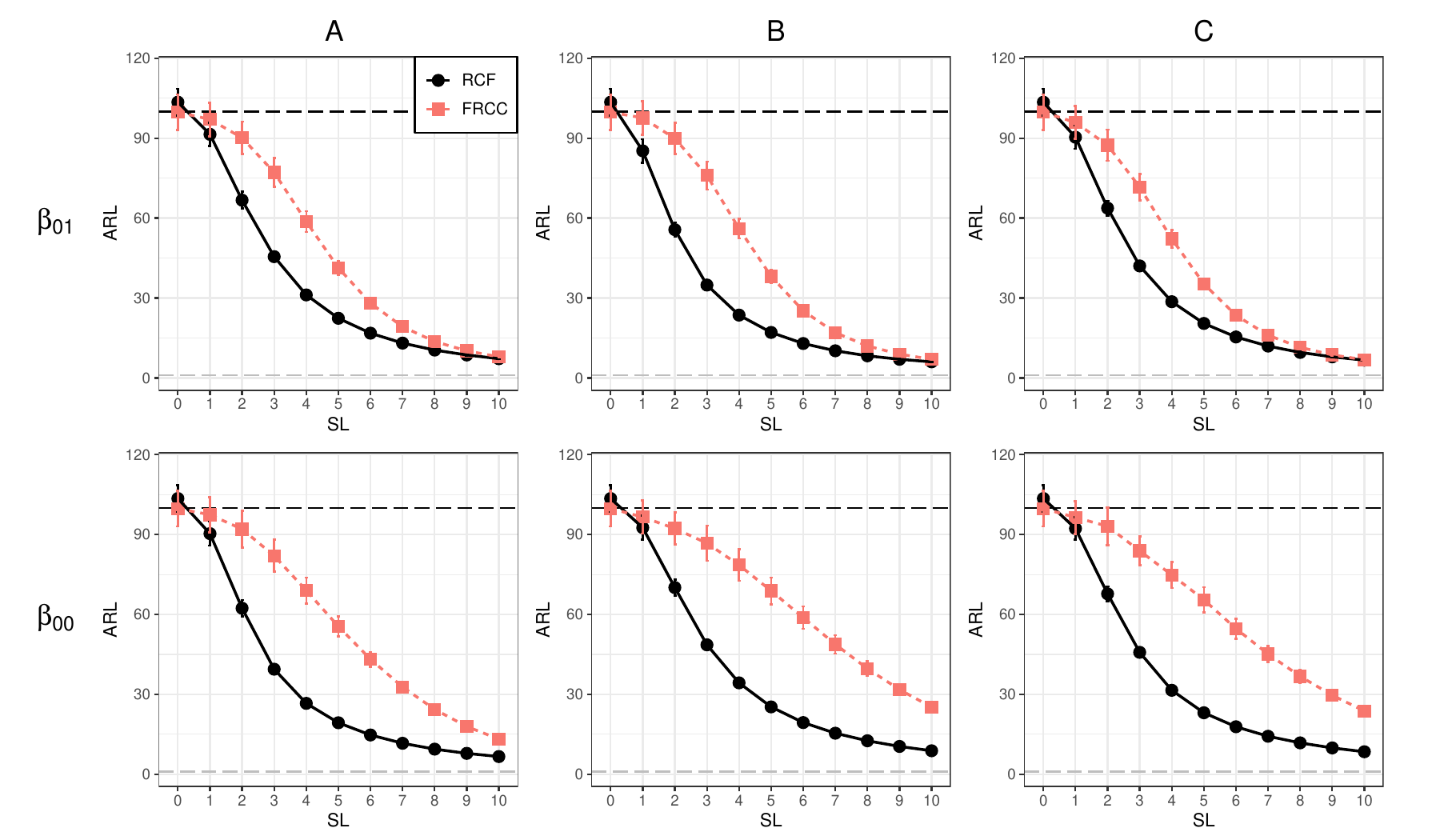}
    \caption{Mean Phase II $ARL$ over Monte Carlo simulation runs and corresponding $95\%$ confidence interval achieved by RCF and FRCC as a function of the severity level SL, for $p = 3$ functional covariates.}
    \label{fig:p_3}
\end{figure}
\begin{figure}
    \centering
    \includegraphics[width=1\linewidth]{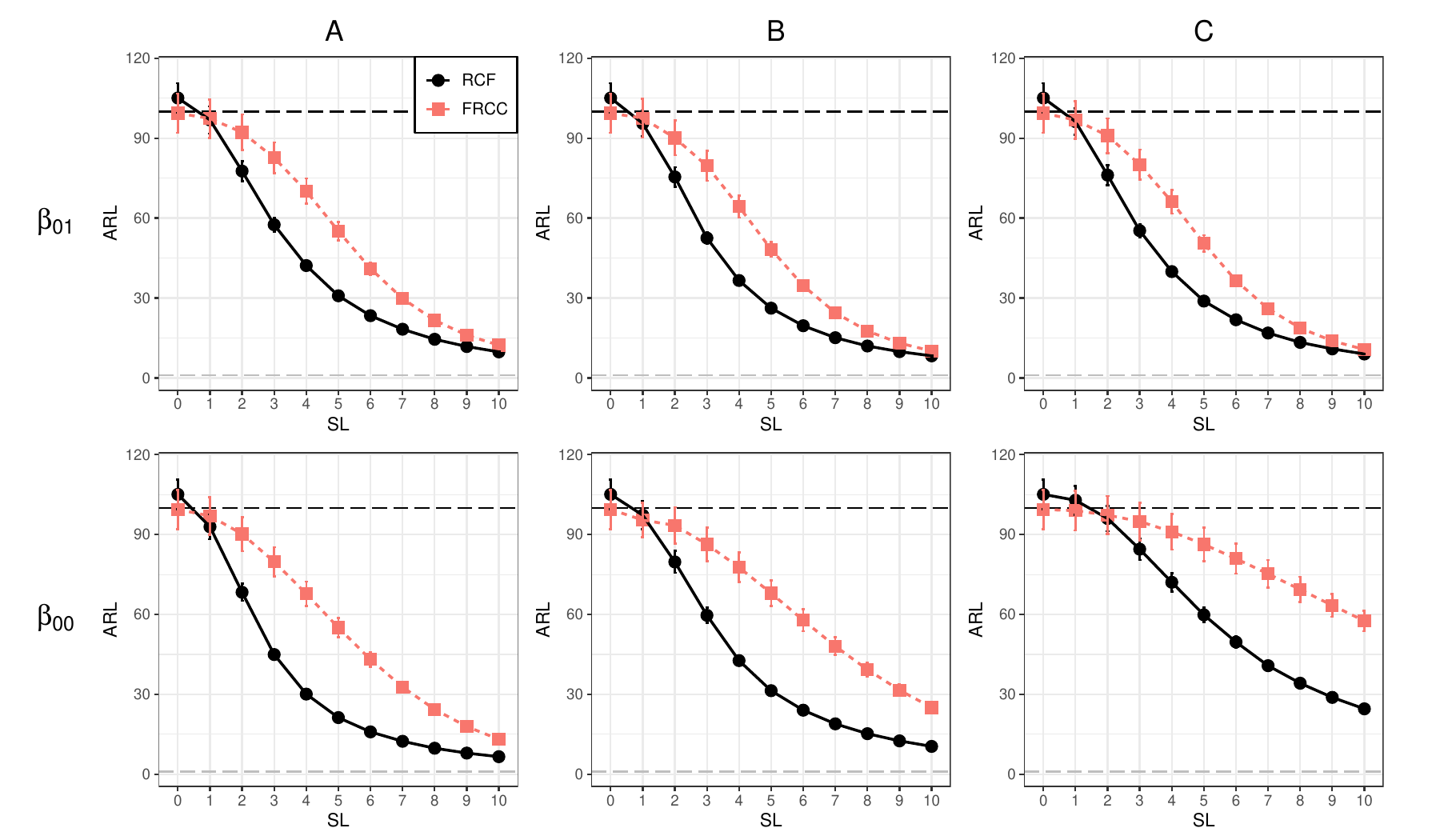}
    \caption{Mean Phase II $ARL$ over Monte Carlo simulation runs and corresponding $95\%$ confidence interval achieved by RCF and FRCC as a function of the severity level SL, for $p = 5$ functional covariates.}
    \label{fig:p_5}
\end{figure}

Simulation results for $p = 1,3,5$ are depicted in Figure \ref{fig:p_1}, Figure \ref{fig:p_3}, and Figure \ref{fig:p_5}, respectively, where the mean $ARL$ over the Monte Carlo simulation runs, together with the corresponding $95\%$ confidence interval, are plotted as a function of $SL$ for each method and shift type. In these figures, the first line of panels refers to shifts in the regression coefficient function $\beta_{01}(s,t)$, while the second line refers to the functional intercept $\beta_{00}(t)$.
In the first scenario ($p = 1$), the proposed RCF control chart outperforms FRCC and AUE (with $\gamma =-0.5)$ for every combination of shift types, both in the case of a shift in the regression coefficient function $\beta_{01}(s,t)$ and in the functional intercept $\beta_{00}(t)$. The performance gap between the proposed method and FRCC is larger at small and intermediate $SL$s, but narrows as $SL$ increases. This is expected, since the proposed RCF control chart accumulates observations to estimate the current model parameters through the exponentially weighted moving loss. At the same time, FRCC uses only the current observation to signal an OC state. On the other hand, the performance gap between the proposed method and AUE (with $\gamma=-0.5$) increases with $SL$.
Conversely, AUE (with $\gamma = 0.25$) performs comparably to the RCF control chart under a shift in the regression coefficient function $\beta_{01}(s,t)$. 
At the same time, it is outperformed under a shift in the intercept function $\beta_{00}(t)$. This last result is also expected because AUE does not directly consider the intercept function, and such a shift can be detected only indirectly through its effect on the response, rather than through a specific model component.
In the second and third scenarios ($p = 3$ and $p = 5$), the results follow the same qualitative pattern observed for $p = 1$, with the proposed RCF control chart outperforming FRCC for every combination of shift type, both for a shift in the regression coefficient function $\beta_{01}(s,t)$ and in the intercept function $\beta_{00}(t)$. As in the case $p = 1$, the performance gap is larger at small and intermediate $SL$s and narrows as $SL$ increases, consistent with the accumulation of observations through the exponentially weighted moving loss in the proposed method, whereas FRCC relies only on the current observation. Moreover, as the number of functional covariates $p$ increases, the performance of both methods  decreases slightly regardless of $SL$, while their relative behavior remains unchanged.
Under the IC condition, all methods achieve $ARL$ values close to the prespecified $ARL_0$ (100), demonstrating their ability to control false alarms.
Finally, we assess the post-signal diagnostic procedure in Section \ref{post-signal} using the F1 score to identify the component responsible for the OC condition. The procedure identifies the shifted component reliably for $p = 1$, with consistently high F1 scores for shifts in the regression coefficient function. 
Its accuracy decreases as the number of functional covariates increases, reflecting the difficulty of attributing the change among a larger number of components. For $p = 3, 5$, it also generally depends non-monotonically on shift severity, being lowest at the smallest severity levels, peaking at moderate ones, and settling at modest values at the largest ones.
Supplementary Materials~\ref{sec_supp_diag} contains the complete results and a more detailed discussion.

\section{Case Study}
\label{sec_real}
\begin{figure}
    \centering
    \includegraphics[width=.75\linewidth]{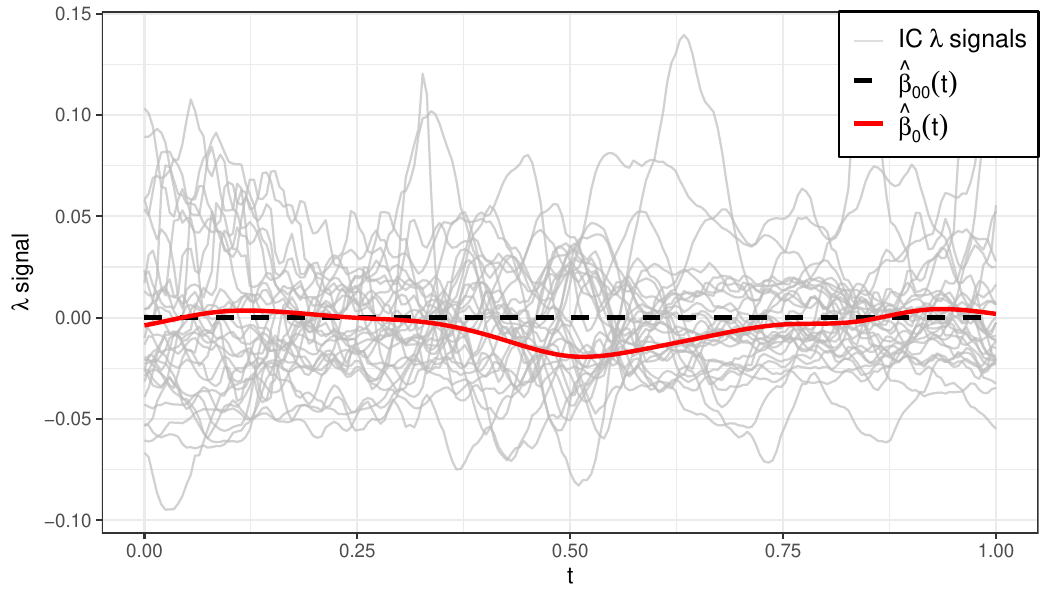}
    \caption{IC intercept function $\hat{\beta}_{00}(t)$ (dashed line) of the centered IC $\lambda$ signals and the estimated OC intercept function $\hat{\beta}_{0}(t)$ (solid line) at the first OC alarm issued by the proposed RCF control chart on top of a sample of 30 centered IC functional response observations (thin line).}
    \label{fig:int_0C_cs}
\end{figure}

To illustrate the practical applicability of the proposed RCF control chart, we present a case study on monitoring the $\lambda$ signal in an automotive after-treatment system.
This application has been previously considered by \cite{gahrooei2019process} and \cite{gomez2020functional}, using the engine data described in \cite{pacella2018unsupervised}.
\begin{figure}
    \centering
    \includegraphics[width=1\linewidth]{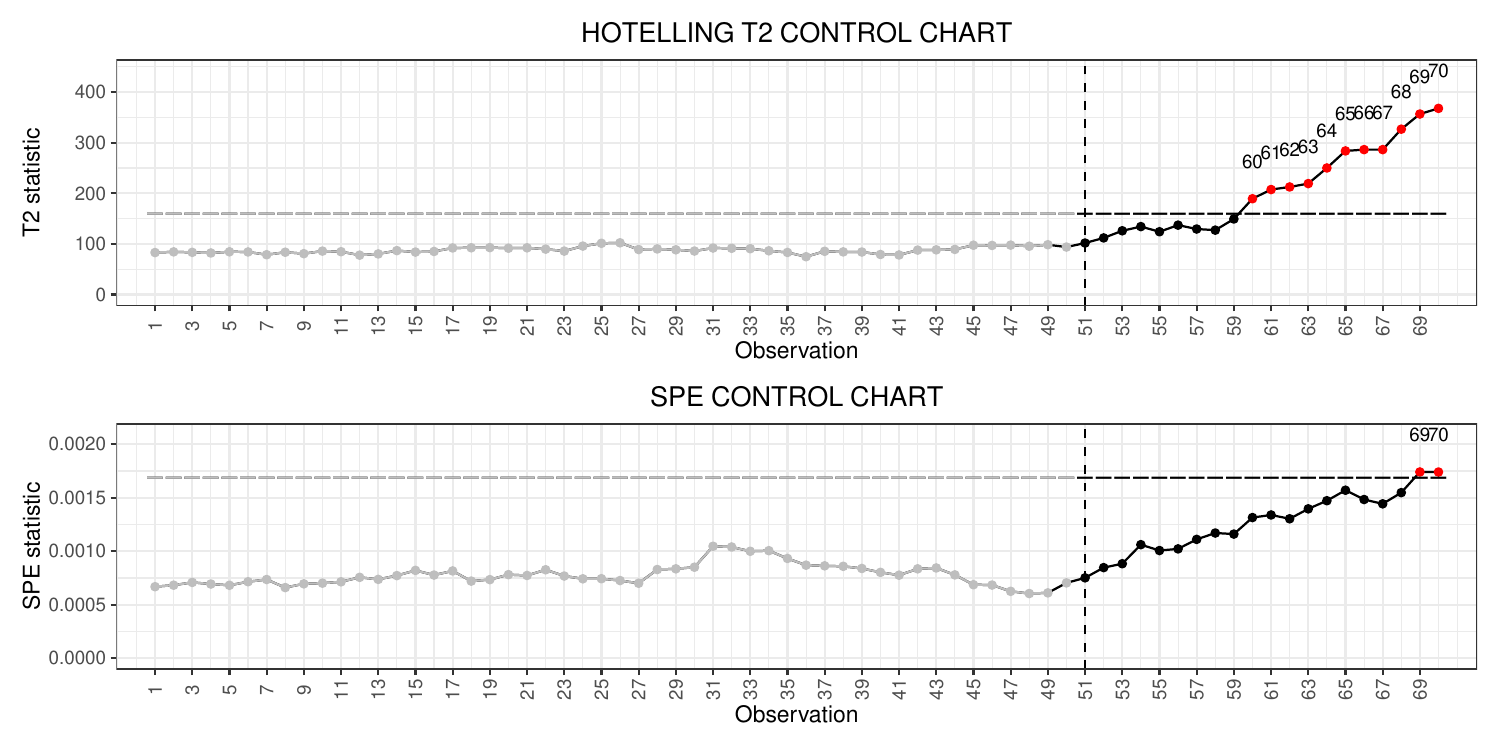}
    \caption{The proposed RCF control chart applied to the OC sequence data in the case study. The grey dots on the left of the dashed vertical line correspond to 50 IC observations, while the black and red dots correspond to OC observations. The RCF control chart signaled at observation 60, which corresponds to an $RL = 10$.}
    \label{fig:cc}
\end{figure}
Modern diesel engines reduce $\mathrm{NO_x}$ emissions through an NSC, in which the exhaust gas is treated in two alternating phases: an adsorption phase, during which $\mathrm{NO_x}$ molecules are trapped by an adsorber, and a regeneration phase, during which the stored $\mathrm{NO_x}$ is catalytically reduced. The regeneration phase lasts 30 to 90 seconds and requires maintaining combustion in a rich air-to-fuel condition, monitored via the $\lambda$ signal measured upstream of the NSC. During correct regeneration, $\lambda$ is expected to remain within a narrow set-point interval, while a $\lambda$ value below an acceptability threshold, known as a $\lambda$-undershoot fault, degrades the catalyst's conversion efficiency and increases emissions \citep{pacella2018unsupervised}. While monitoring the $\lambda$ sensor readily detects such a fault, its root causes are not uniquely determined and may stem from changes in the relationship between $\lambda$ and several engine state variables \citep{gomez2020functional}.
Monitoring the stability of the functional relationship between the $\lambda$ signal and the engine state variables over time is therefore important in practice. A change in this relationship may indeed indicate a fault or a deviation in the after-treatment process that compromises emission control.
The $\lambda$ signal, measured upstream of the NSC, is thus taken as the functional response, and a set of five engine state variables as functional covariates, which are known to directly influence the $\lambda$ signal \citep{gahrooei2019process}, namely the air aspirated per cylinder, the engine rotational speed, the injection quantity, the boost pressure, and the inner torque.
The data consist of real engine observations of the $\lambda$ signal and the engine state variables during the $\lambda$-undershoot event, each recorded over a two-second interval on a grid of $203$ points. The OC observations therefore correspond to events in which $\lambda$ falls below the acceptability threshold. In contrast, IC observations, which represent correct regeneration, are obtained by replacing the portion of the domain affected by the $\lambda$-undershoot with values generated from a model fitted to the non-undershoot portion of the real data, where $\lambda$ remains within its set-point interval.
Following Section \ref{implementation}, we partition the available IC observations into a training set and a tuning set, comprising $99$ and $100$ functional observations, respectively. We use the training set to estimate the IC coefficient vector, select the roughness penalty parameter, and estimate the MFPCA model. The tuning set is used to compute the control limits, with $n_{\mathrm{seq}} = 200$ sequences of length $l_{\mathrm{seq}} = 300$ generated to attain a prespecified $ARL_0 = 100$. We set the number of B-spline bases to $15$ for both the functional response and the functional covariates, and the fraction of variance explained to $0.95$. We set the forgetting factor to $\eta = 0.95$. We use a transient of $200$ IC observations to reach steady state before monitoring.
\begin{figure}
    \centering
    \includegraphics[width=0.5\linewidth]{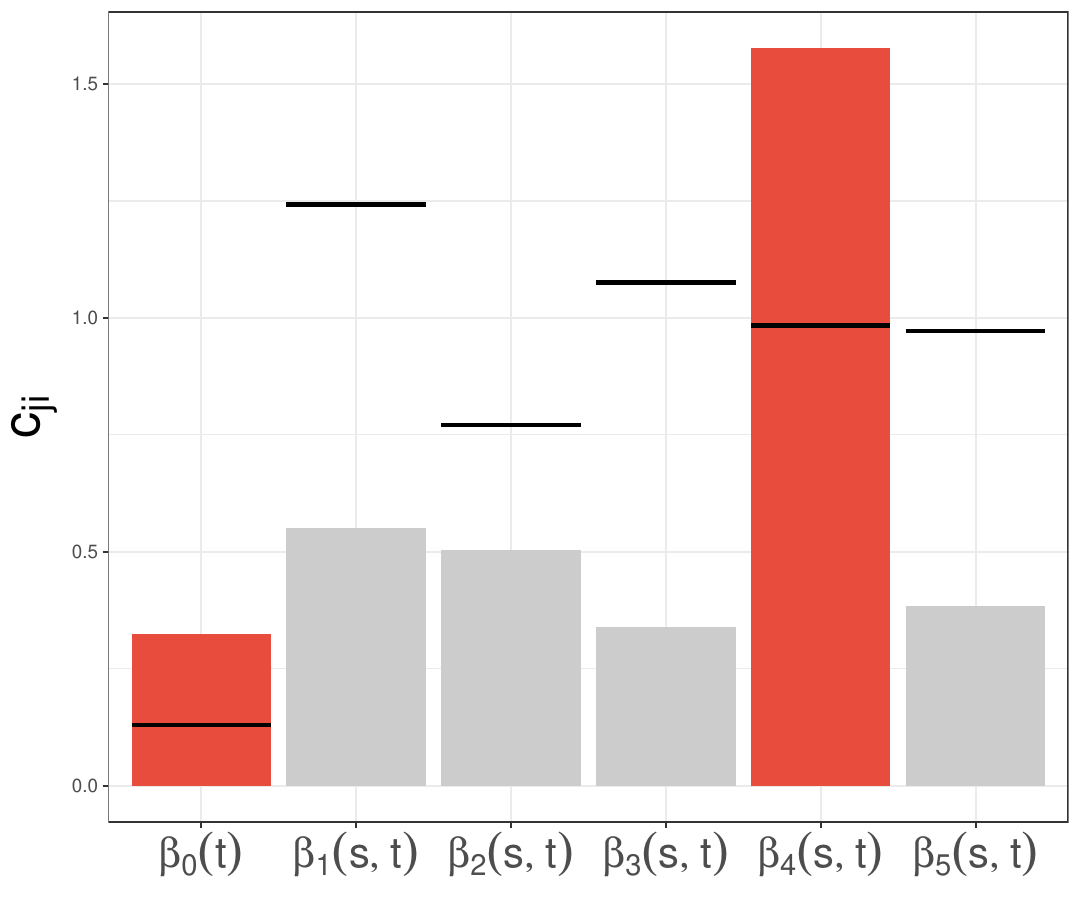}
    \caption{Contribution plot corresponding to the Hotelling $T^2$ statistic of the proposed RCF control chart. A shift in the intercept function $\beta_0(t)$ and in the regression coefficient function $\beta_4(s,t)$ corresponding to the relationship between the $\lambda$ signal and boost pressure is identified.}
    \label{fig:contr}
\end{figure}
We apply the proposed RCF control chart to a randomly sampled OC sequence of 20 observations. Figure \ref{fig:cc} shows the $T^2$ and $SPE$ control charts of the proposed method. Grey points correspond to the last 50 IC observations used to reach the steady state and are separated from the OC sequence by the dashed vertical lines, while black and red points correspond to OC observations. The dashed horizontal lines represent the control limits. The Hotelling $T^2$ of the proposed method signals an OC condition at observation 60, corresponding to a run length ($RL$) of 10, and rises steadily up to the last OC observation.
The contribution plot for the $T^2$ statistic at the signal, shown in Figure \ref{fig:contr}, indicates that the components responsible for the OC condition are the intercept function and the regression coefficient function associated with boost pressure.
Figure \ref{fig:int_0C_cs} shows the estimated intercept function at the signal (solid line), which exhibits a marked decay in the central part of the domain relative to its IC value. This is consistent with the physical signature of the $\lambda$-undershoot fault, which manifests as a drop in the $\lambda$ level during the regeneration window, and confirms that the diagnostic attributes the detected change to the component corresponding to the known fault.
\begin{figure}
    \centering
    \includegraphics[width=1\linewidth]{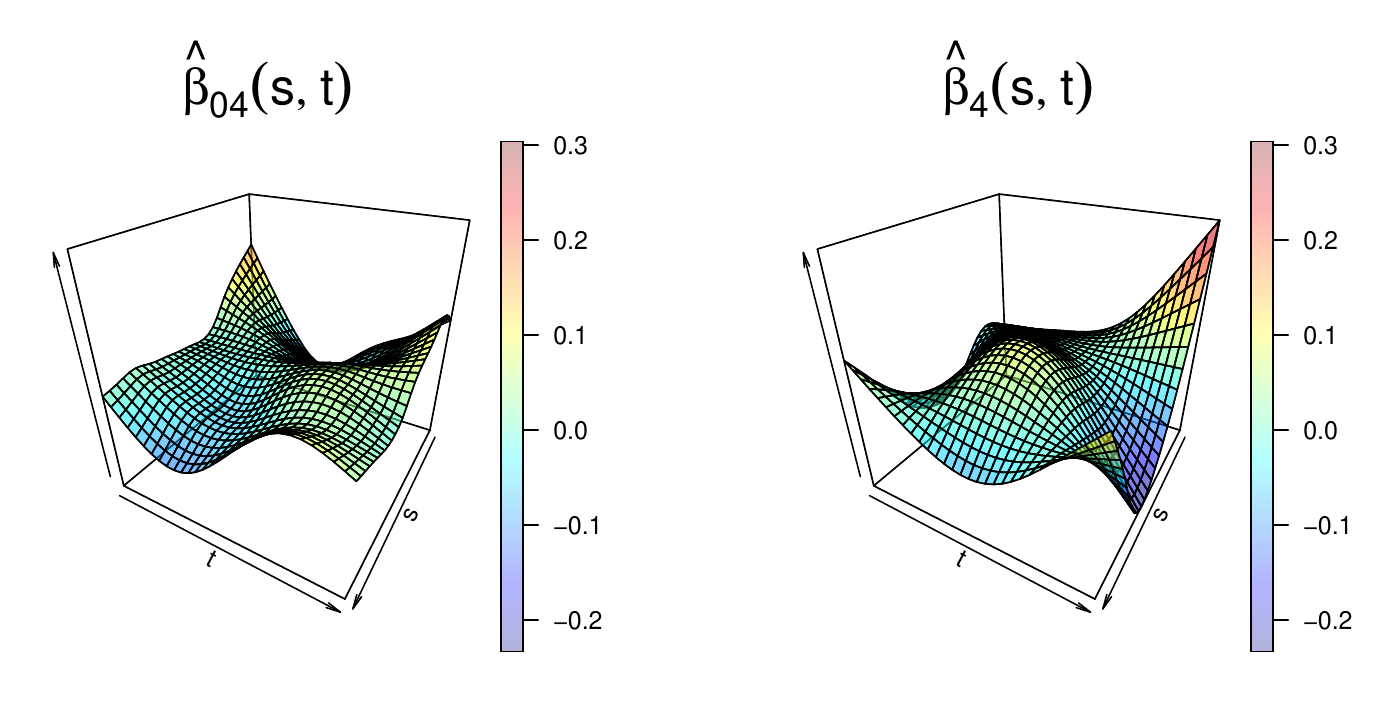}
    \caption{Estimated IC regression coefficient function $\hat{\beta}_{04}(s,t)$ (left) and the estimated OC regression coefficient function $\hat{\beta}_{4}(s,t)$ (right) signaled by the proposed RCF control chart corresponding to the boost pressure}
    \label{fig:beta4_0C_cs}
\end{figure}
The regression coefficient function associated with the centered boost pressure is also flagged, and its estimate, $\hat{\beta}_4(s,t)$, together with its IC counterpart, $\hat{\beta}_{04}(s,t)$, is depicted in Figure \ref{fig:beta4_0C_cs}.
Compared to the IC $\hat{\beta}_{04}(s,t)$, the OC estimate $\hat{\beta}_4(s,t)$ is overall less flat, especially in the last part of the $t$ domain, where it takes a parabolic shape and shows a small positive bump in its central region. Two features stand out. First, in the last part of the $t$ domain, the $\lambda$ signal is more strongly associated with the centered boost pressure, positively over the first part of the $s$ domain and negatively over the central part. Second, in the central part of the $t$ domain, the $\lambda$ signal is also more strongly associated with the centered boost pressure, specifically over the central part of the $s$ domain, corresponding to the small positive bump in $\hat{\beta}_4(s,t)$.
Including boost pressure in the OC condition aligns with the root-cause analysis of \cite{gomez2020functional}, which associates the $\lambda$-undershoot with changes in the engine's intake and operating dynamics during the fault.
For comparison, we apply the FRCC to the same OC sequence. Its $T^2$ statistic shows a small but visible upward shift over the OC observation region relative to the IC one, indicating that it responds, albeit minimally, to the change. 
However, because it monitors each observation individually, it is insensitive to the small sustained shift induced by the fault. Indeed, the corresponding $T^2$ statistic exceeds the control limit only at a single isolated observation, while its $SPE$ statistic never exceeds it (see Supplementary Materials~\ref{supp_frcc}). 
In contrast, the proposed RCF control chart accumulates information across observations via an exponentially weighted moving loss and detects the small persistent shift that causes both monitoring statistics to increase steadily beyond their control limits.
Moreover, even when FRCC responds to the change, its statistics are computed on the functional residuals, so a signal would locate the change in the domain of the $\lambda$ signal. In contrast, the diagnostics in Figure~\ref{fig:contr} attribute it to the intercept function and to the coefficient function associated with the boost pressure, and Figure~\ref{fig:beta4_0C_cs} shows how the latter has changed.

\section{Conclusions}
\label{sec_conclusions}
In this article, we proposed the regression coefficient function (RCF) control chart for the sequential monitoring of the functional linear regression relationship between a functional response and multiple functional covariates. At each sampling stage, we update the regression coefficient functions and the intercept function using an exponentially weighted moving loss that accumulates information across observations while progressively discounting past observations. We monitor the resulting estimates using Hotelling's $T^2$ and squared prediction error ($SPE$) control charts, which monitor variation within the retained principal component subspace and in its orthogonal residual subspace. A post-signal diagnostic procedure, based on the decomposition of the monitoring statistics into contributions from individual components, also identifies which regression coefficient functions or the intercept function are responsible for a detected change.
An extensive Monte Carlo simulation study showed that the proposed method outperforms the functional regression control chart of \cite{centofanti2021bfunctional} across all the considered scenarios, shift types, and numbers of functional covariates. 
The advantage arises from accumulating information across observations, rather than testing each one alone against a model held fixed from Phase I, and from considering the coefficient functions directly, which is what allows an OC signal to
be attributed to a specific model component.
Compared with the method of \cite{aue2014dependent}, which is restricted to a single functional covariate, the proposed method attains comparable performance for shifts in the regression coefficient function and superior performance for shifts in the intercept function, while remaining applicable to an arbitrary number of covariates and requiring no choice of a change-location sensitivity parameter. A case study on monitoring the $\lambda$ signal in an automotive after-treatment system demonstrated the method's practical applicability, with post-signal diagnostics attributing the detected $\lambda$-undershoot fault to the components corresponding to its known physical signature.
As anticipated in the Introduction, the sequential estimates are themselves interpretable and able to describe how the relationship may have changed, as illustrated in the case study.
These benefits come at the cost of a recursive state, which requires a transient before the steady state is attained, and of reduced advantage against large, abrupt changes, for which a single observation already provides sufficient evidence.
Several directions remain open for future research. While the present work focuses on the functional linear model, extending the proposed method to nonlinear functional relationships is a natural and practically relevant development, as is its extension to settings with serially dependent observations, particularly in applications where the process evolves. 

\section*{Supplementary Materials}
The Supplementary Materials contain additional details on data generation, simulation results, and the \textsf{R} code to reproduce graphs and results in the simulation study.

\section*{Disclosure of Interest}
The authors declare no conflicts of interest.

\section*{Data Availability Statement}
The simulated data supporting the findings of the simulation study can be reproduced using the \textsf{R} code provided in the Supplementary Materials. The engine data used in the case study are described in \cite{gomez2020functional} and are available from the corresponding author upon reasonable request.

\if0\blind{
\section*{Acknowledgments}
The computing resources and related technical support used for this work have been provided by the CRESCO/ENEAGRID High Performance Computing infrastructure and its staff \citep{9188135}. CRESCO/ENEAGRID High Performance Computing infrastructure is funded by ENEA, the Italian National Agency for New Technologies, Energy and Sustainable Economic Development, and by Italian and European research programmes, see \url{http://www.cresco.enea.it/english} for information.

\section*{Funding}
The work of Davide Forcina was funded, and the work of Kamran Paynabar was partially funded, by the Department of Industrial Engineering, University of Naples Federico II, within the project SMADI -- Statistical Methods for Anomaly Detection in Industry 4.0.
}\fi

\bibliographystyle{apalike}
\setlength{\bibsep}{5pt plus 0.2ex}
{\small
\spacingset{1}
\bibliography{ref}
}

% =====================================================================
% Supplementary Materials -- body only.
% Included at the end of main.tex via \input{Supplementary_body}.
% No preamble, no \begin{document}/\end{document}, no bibliography:
% citations here go into the main bibliography.
% =====================================================================

\clearpage
\spacingset{1}
\begin{center}
  {\LARGE\bf Supplementary Materials to ``Statistical Process Monitoring of Functional Linear Models''}
\end{center}
\medskip
\spacingset{1.45} % DON'T change the spacing!

\appendix
\numberwithin{equation}{section}
\setcounter{figure}{0}\renewcommand{\thefigure}{S\arabic{figure}}
\setcounter{table}{0}\renewcommand{\thetable}{S\arabic{table}}

\section{Details on data generation in the simulation study}
\label{supp_datagen}
The data are generated according to the FLM
\begin{equation}
    \label{flm_sim}
    Y(t) = \sum_{j = 1}^p\int_{[0,1]}\beta_{0j}(s,t)X_j(s)ds + \varepsilon(t), \quad t \in [0,1],
\end{equation}
where the number of covariates $p = 1, 3, 5$ defines different scenarios.
The functional covariates $X_j(s)$, $j = 1,\dots,p$, are generated from the eigenfunctions $\lbrace\varphi_l\rbrace_{l=1,2,\dots}$ obtained from the spectral decomposition of pre-defined correlation functions $G(s,t)$ through the following steps.
\begin{enumerate}
    \item Set $G(s,t) = G(z)$, where $z = |s-t|$, as one of the correlation functions listed in Table \ref{ta_corf}.
    \item Calculate the eigenvalues $\kappa_l$ and the corresponding eigenfunctions $\varphi_l$, $l = 1, 2, \dots, L^*$, of $G$, where $L^*$ is set equal to $20$.
    \item Define $X_j$ as
    \begin{equation*}
    X_j = \sum_{l=1}^{L^*}a_l\varphi_l,
    \end{equation*}
    with $\bm{a}_{L^{*}} = \left(a_1,\dots,a_{L^{*}}\right)^{T}$ generated from a multivariate normal distribution with zero mean and covariance $\Cov\left(\bm{a}_{L^{*}}\right) = \diag\left(\kappa_1,\dots,\kappa_{L^{*}}\right)$.
\end{enumerate}
The correlation functions are assigned to the covariates cyclically, so that $X_1$ and $X_4$ use the powered exponential, $X_2$ and $X_5$ the Bessel, and $X_3$ the Gaussian correlation function. Correlation functions govern the smoothness of the functional covariates, whereas the regression coefficient functions $\beta_{0j}(s,t)$ are generated separately. We generate the covariates using the \texttt{R} package \texttt{funcharts} \citep{capezza2023funcharts}.
\begin{table}
\caption{Correlation functions and parameters for data generation in the simulation study.}
\label{ta_corf}
\centering
\resizebox{0.75\linewidth}{!}{
\begin{tabular}{lccc}
\toprule
Type & & $\rho$ & $\nu$\\
\midrule
Bessel \citep{abramowitz1948handbook} &
$G_{\mathrm{B}}(z)=
\left(\frac{|z|}{2\rho}\right)^{\nu}
\sum_{j=0}^{\infty}
\frac{\left[-\left(\frac{|z|}{2\rho}\right)^2\right]^j}
{j!\,\Gamma(\nu+j+1)}$
& 0.25 & 0 \\[0.25cm]

Gaussian \citep{abrahamsen1997review} &
$G_{\mathrm{G}}(z)=\exp\!\left[-\left(\frac{|z|}{\rho}\right)^2\right]$
& 1 & -- \\[0.25cm]

Powered exponential \citep{stein1999interpolation} &
$G_{\mathrm{PE}}(z)=\exp\!\left[-\left(\frac{|z|}{\rho}\right)^\nu\right]$
& 1 & 0.5 \\
\bottomrule
\end{tabular}}
\end{table}

The regression coefficient functions $\beta_{0j}(s,t)$, $j = 1,\dots,p$, are generated as smooth bivariate surfaces over $[0,1]\times[0,1]$, each drawn independently from a two-dimensional Gaussian process with zero mean and separable squared-exponential covariance function
\begin{equation*}
K\big((s,t),(s',t')\big) = \sigma^2 \exp\!\left(-\frac{(s-s')^2}{2\ell_s^2}\right)\exp\!\left(-\frac{(t-t')^2}{2\ell_t^2}\right),
\end{equation*}
with marginal variance $\sigma^2 = 1$ and length scales $\ell_s = \ell_t = 1/3$, which yield smooth surfaces with a few gentle oscillations across the domain. We draw the surfaces once using a fixed random seed and hold them fixed across all Monte Carlo runs, so they represent the fixed IC relationship, while the functional covariates, the error term, and the shifts are regenerated at each run. For example, Figure~\ref{fig:betas} of the main article depicts the coefficient functions $\beta_{0j}$ corresponding to the first three functional covariates. We set the IC intercept function to zero.
Moreover, to obtain noisy realizations of the response, the error term in Equation~\eqref{flm_sim} is generated as $\varepsilon(t) = h(t)\tilde{\varepsilon}(t)$, where $\tilde{\varepsilon}(t) = \sum_{k=1}^{20} e_k\psi_k^e(t)$, the $e_k$ are independent standard normal random variables, and the $\psi_k^e(t)$ are cubic B-splines with an evenly spaced knot sequence. The function $h(t)$ is chosen so that, at each $t$, a prescribed signal-to-noise ratio (SNR) is attained, that is, $\mathrm{Var}(Y^*(t))/\mathrm{Var}(\varepsilon(t)) = \mathrm{SNR}$, where $Y^*(t) = \sum_{j=1}^p\int_{[0,1]}\beta_{0j}(s,t)X_j(s)ds$ is the noiseless response, which gives
\begin{equation*}
h(t) = \sqrt{\frac{\mathrm{Var}(Y^*(t))}{\mathrm{SNR}\cdot\mathrm{Var}(\tilde{\varepsilon}(t))}}.
\end{equation*}
We set the SNR to $10$. We assume the generated data are observed discretely at $100$ equally spaced points over $[0,1]$, for both the functional response and the functional covariates.

To assess the performance of the RCF control chart under OC conditions, we consider three types of shifts, namely shift A, shift B, and shift C, which are applied separately to the regression coefficient function $\beta_{01}(s,t)$ and to the intercept function $\beta_{00}(t)$, whose IC value is zero. More specifically, the shifts are generated by adding to $\beta_{01}(s,t)$ and $\beta_{00}(t)$ the following quantities, containing a constant $d$ that defines the severity level ($SL$) of the shift:
\begin{itemize}
    \item Shift A is obtained by adding
    \begin{equation*}
    \delta_A(s) = \sqrt{p}\,c_A \left[\frac{d\,0.07}{(0.75-0.5)^2}(s-0.5)^2 - d\,0.07\right]\mathbf{1}_{[0.25,0.75]}(s),
    \end{equation*}
    which produces a localized decrease in the central part of the domain;
    \item Shift B is obtained by adding
    \begin{equation*}
    \delta_B(s) = \sqrt{p}\,c_B\, d\,0.05\,\sin(2\pi s),
    \end{equation*}
    which produces a sinusoidal shape;
    \item Shift C is obtained by adding
    \begin{equation*}
    \delta_C(s) = -\sqrt{p}\,c_C\, d\,s^2,
    \end{equation*}
    which produces a quadratic decrease.
\end{itemize}
The intercept function $\beta_{00}(t)$ is shifted by the same quantities, with $s$ replaced by $t$.
The factor $\sqrt{p}$ keeps the $SL$ comparable as the number of covariates grows, and $c_A$, $c_B$, $c_C$ are shift-specific scale constants equal to $10$, $20$, and $0.7$, respectively, for the coefficient function, and to $90$, $60$, and $2.5$ for the intercept function. For each shift, the severity constant takes the values $d = 0.03\, SL$, $SL = 0,\dots,10$, where $SL = 0$ corresponds to the IC state.

\section{Post-signal diagnostic performance}
\label{sec_supp_diag}
To evaluate the ability of the post-signal diagnostic procedure of Section~\ref{post-signal} to correctly identify the component responsible for an OC condition, we use the same scenarios ($p = 1, 3, 5$), shift types (A, B, C), and $SL$s as in the simulation study of Section~\ref{sec_sim}. In particular, we place a single change point at the beginning of the monitoring sequence, immediately after the transient of $200$ observations used to reach steady state. In each run, either the regression coefficient function $\beta_{01}(s,t)$ or the intercept function $\beta_{00}(t)$ is shifted, only.

For each Monte Carlo run, the post-signal diagnostic starts at the first alarm of either $T^2$ or $SPE$ control charts, without waiting for further observations.
Consistent with Section~\ref{post-signal}, we examine the contributions of the statistic responsible for the signal, that is, the $T^2$ contributions $c^{T^2}_{ji}$ if the signal is issued by the $T^2$ statistic, and/or the $SPE$ contributions $c^{SPE}_{ji}$ if it is issued by the $SPE$ statistic.
We flag component $j \in \{0, 1, \dots, p\}$ as responsible when its contribution exceeds the corresponding contribution limit. We classify the flagged components as true positives (TP), false positives (FP), and false negatives (FN), and compute the F1 score as $\mathrm{F1} = 2\mathrm{TP}/(2\mathrm{TP}+\mathrm{FP}+\mathrm{FN})$. The per-run F1 scores are then averaged over the $100$ Monte Carlo runs. Note that the diagnostic procedure does not apply under the IC state ($SL=0$). As the number of covariates $p$ increases, the number of components among which the shifted one must be identified grows, so that more components can be falsely flagged.

\begin{figure}
    \centering
    \includegraphics[width=1\linewidth]{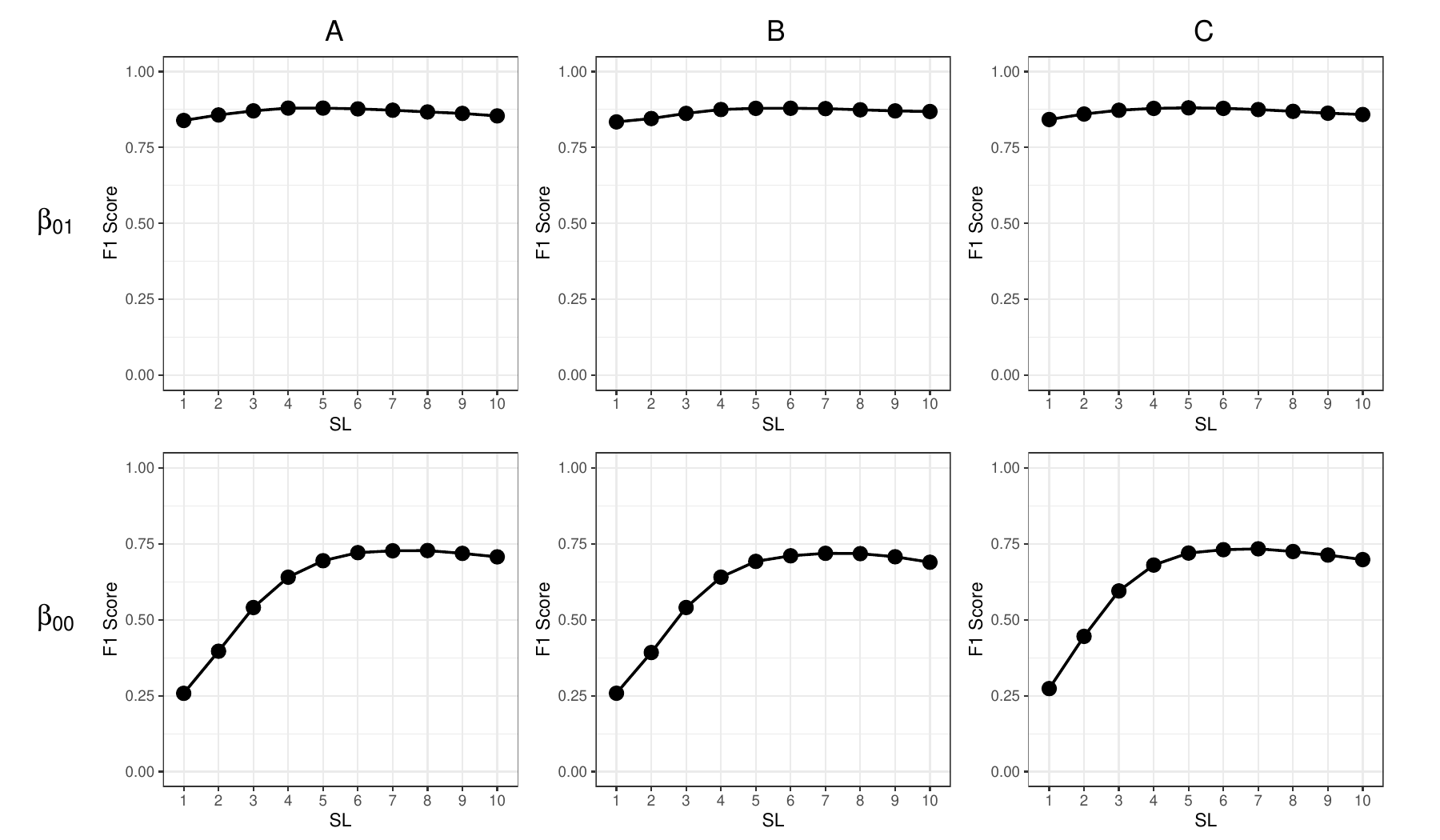}
    \caption{Mean F1 score achieved in Phase II by the RCF control chart post-signal diagnostic for $p = 1$ functional covariate as a function of $SL$.}
    \label{fig:p_1_f1}
\end{figure}

\begin{figure}
    \centering
    \includegraphics[width=1\linewidth]{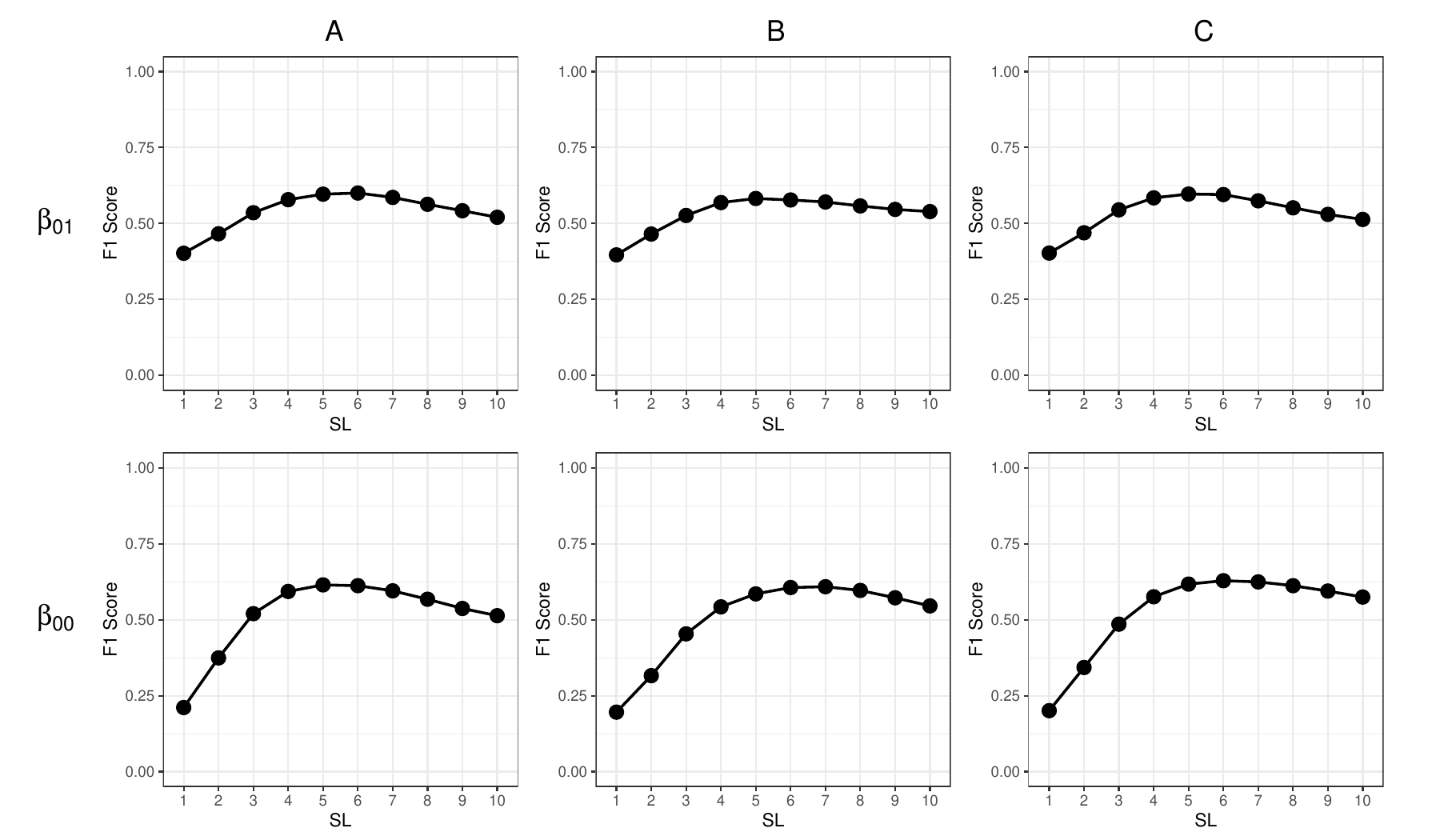}
    \caption{Mean F1 score achieved in Phase II by the RCF control chart post-signal diagnostic for $p = 3$ functional covariates as a function of $SL$.}
    \label{fig:p_3_f1}
\end{figure}

\begin{figure}
    \centering
    \includegraphics[width=1\linewidth]{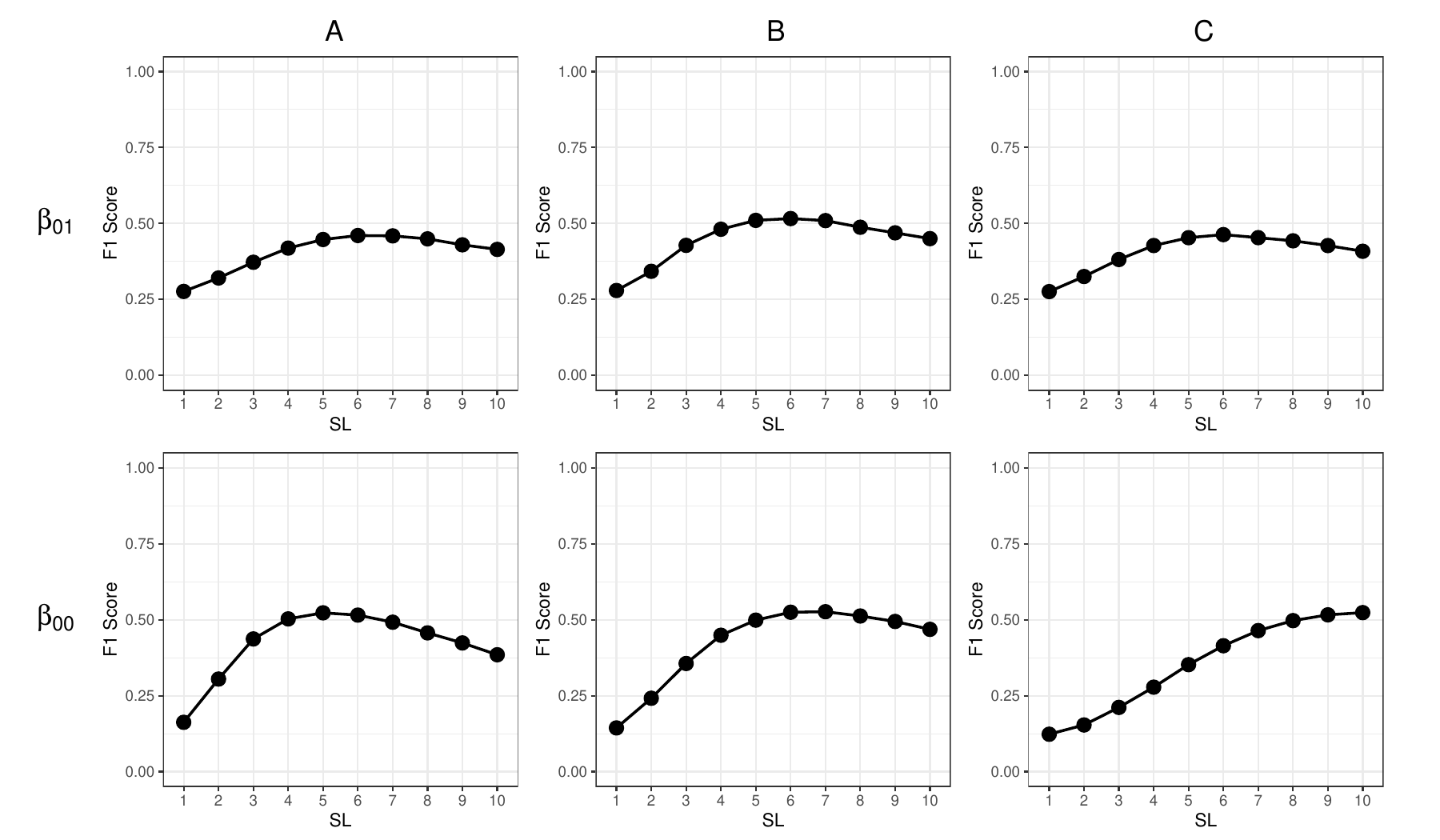}
    \caption{Mean F1 score achieved in Phase II by the RCF control chart post-signal diagnostic for $p = 5$ functional covariates as a function of $SL$.}
    \label{fig:p_5_f1}
\end{figure}

Figure \ref{fig:p_1_f1}, Figure \ref{fig:p_3_f1}, and Figure \ref{fig:p_5_f1} report the average F1 score as a function of $SL$ for $p = 1, 3, 5$, respectively. Each figure comprises six panels arranged in three columns, one per shift type (A, B, C), and two rows, corresponding to a shift in the regression coefficient function $\beta_{01}(s,t)$ (top row) and in the intercept function $\beta_{00}(t)$ (bottom row).
For $p = 1$, the proposed post-signal diagnostic can identify the shifted component. For shifts in the regression coefficient function $\beta_{01}(s,t)$, the average F1 score is high, around $0.85$, and essentially constant across $SL$s for all three shift types. For shifts in the intercept function, the F1 score starts around $0.25$ at $SL = 1$, increases to about $0.7$ at $SL = 5$, and then remains stable.
For $p = 3$ and $p = 5$, the behavior is qualitatively similar for both scenarios, but with lower F1 scores overall and a non-monotone dependence on $SL$. For $p = 3$, and a shift in $\beta_{01}(s,t)$, the F1 score starts around $0.4$ at $SL = 1$, rises to about $0.6$ at $SL = 5$, and then declines gradually to about $0.5$ at the largest $SL$s; for a shift in the intercept function, it starts slightly below $0.25$, rises to about $0.625$ around $SL = 5$--$6$, and then decreases to just above $0.5$. The scores for $p = 5$ follow the same pattern as for $p = 3$, shifted slightly downward.
Two effects therefore govern the diagnostic performance. First, as the number of covariates $p$ increases, the F1 score decreases because the shifted component must be identified among more candidates and more components can be falsely flagged. Second, as the $SL$ increases, for $p = 3,5$ the F1 score first improves and then deteriorates, while for $p = 1$ it is essentially constant for coefficient shifts and levels off after an initial increase for intercept shifts. This non-monotone behavior reflects the interplay between the $SL$ and the number of OC observations available at the signal. At moderate $SL$s, the control chart signals when the exponentially weighted moving loss has accumulated enough information to estimate the OC coefficient or intercept function reliably, and the contributions correctly localize the shifted component. At large $SL$s, although the shift is more pronounced, the run length is short, and the chart signals after only a few OC observations. The sequential estimate has not yet stabilized on the OC relationship, and the resulting contributions produce more false positives, lowering the F1 score.
The only exception is the intercept shift of type C for $p = 5$, whose F1 score keeps increasing up to the largest $SL$. This is consistent with the explanation above, since in this scenario the run length remains comparatively long even at the largest $SL$ (Figure~\ref{fig:p_5} of the main article), so that the sequential estimate has time to stabilize on the OC relationship before the signal.

\section{Sensitivity Analyses}
\label{supp_sens}
The sensitivity of the proposed RCF control chart to its main tuning parameters, namely the number of basis functions, the forgetting factor $\eta$, and the fraction of variance explained (FVE), is studied by varying one parameter at a time while holding all others at the default values used in the simulation study of Section~\ref{sec_sim}. This analysis is carried out for the representative case $p=3$. For each configuration, we calibrate the control limits as described in Section~\ref{implementation} to achieve the same $ARL_0 = 100$.

\subsection{Number of Basis Functions}
\label{supp_nbasis}
\begin{figure}
    \centering
    \includegraphics[width=1\linewidth]{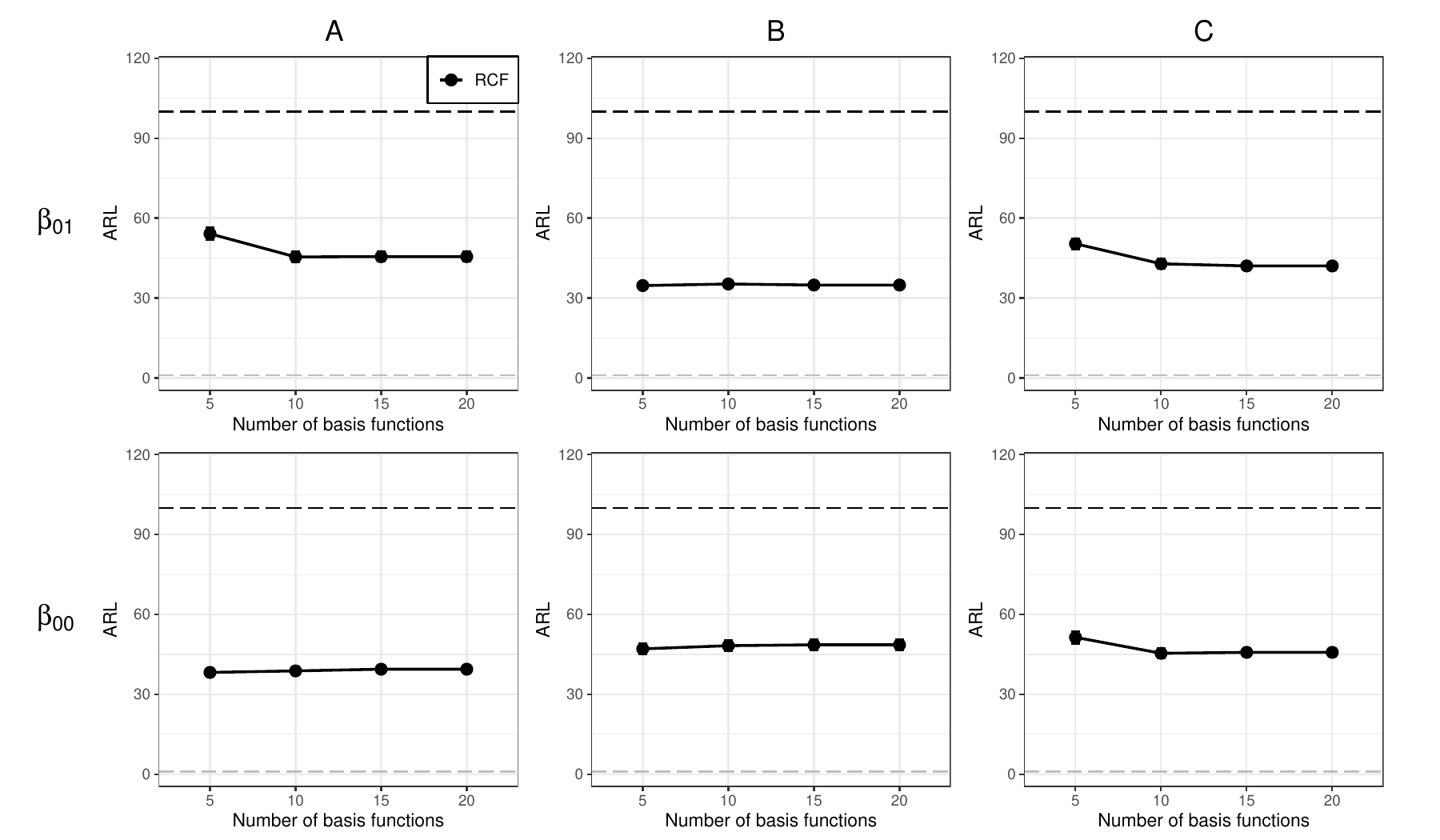}
    \caption{Mean $ARL$ over the Monte Carlo simulation runs, and its corresponding $95\%$ confidence interval achieved in Phase II by RCF, for different values of the number of basis functions $k_y = k_{x_j} = 5, 10, 15, 20$, $j = 1,\dots,p$, for $p = 3$ functional covariates and $SL = 3$.}
    \label{fig:sens_nbasis}
\end{figure}
We vary the number of basis functions used to represent the functional response and covariates over $k_y = k_{x_j} = 5, 10, 15, 20$, $j = 1,\dots,3$, at the fixed severity level $SL = 3$. Figure \ref{fig:sens_nbasis} reports the mean $ARL$ over the Monte Carlo simulation runs, and its corresponding $95\%$ confidence interval for each shift type (columns A, B, C) and shift location (top row, shift in the regression coefficient function $\beta_{01}(s,t)$; bottom row, shift in the intercept function $\beta_{00}(t)$), as a function of the number of basis functions.
Performance degrades only for the smallest value, $k_y = k_{x_j} = 5$, $j = 1,2,3$. For a shift in $\beta_{01}(s,t)$, this occurs under shift types A and C, while the performance under shift type B is unaffected by this behavior. For a shift in the intercept function, the loss occurs only under shift type C. For all values $k_y = k_{x_j} \geq 10$, the mean $ARL$ is essentially unchanged across all shift types and locations. This confirms that the method is practically unaffected by the number of basis functions, provided that it is not too small. It can therefore be set conservatively high to capture the local behavior of the functional response and covariates thanks to the presence of the roughness penalty that controls the smoothness of the estimates.

\subsection{Forgetting factor $\eta$}
\label{supp_eta}
\begin{figure}
    \centering
    \includegraphics[width=1\linewidth]{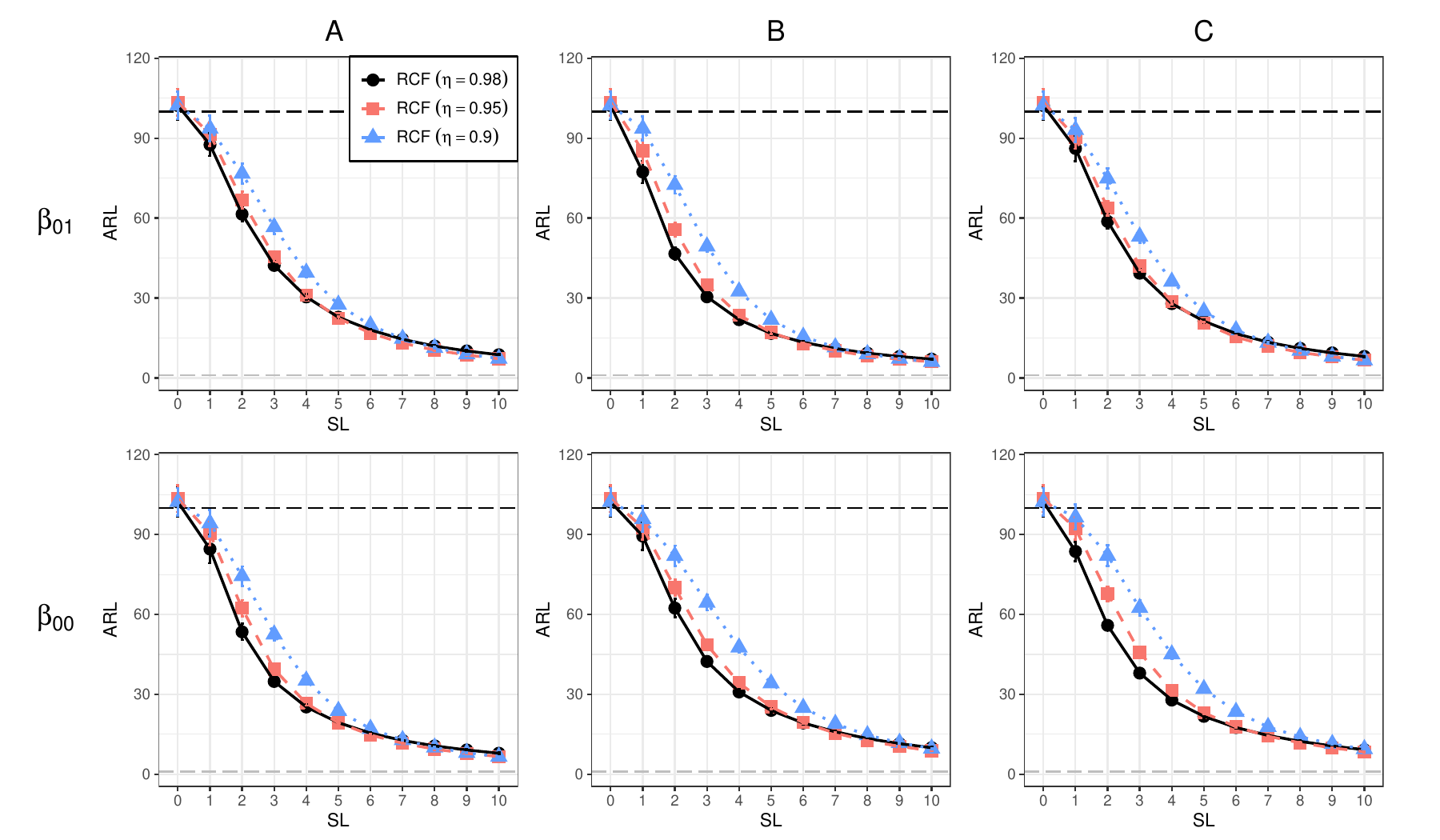}
    \caption{Mean $ARL$ over the Monte Carlo simulation runs, and its corresponding $95\%$ confidence interval achieved in Phase II by RCF, with forgetting factor $\eta = 0.90, 0.95, 0.98$, for $p = 3$ functional covariates as a function of $SL$.}
    \label{fig:sens_eta}
\end{figure}
We vary the forgetting factor over $\eta = 0.90, 0.95, 0.98$, holding all other parameters at their default values. Figure \ref{fig:sens_eta} reports the mean $ARL$ over the Monte Carlo simulation runs, and its corresponding $95\%$ confidence interval as a function of $SL$ for each shift type (columns A, B, C) and shift location in the regression coefficient function $\beta_{01}(s,t)$ (top row), and in the intercept function $\beta_{00}(t)$ (bottom row), where each curve corresponds to a different value of $\eta$.
As discussed in Section~\ref{implementation}, $1-\eta$ plays a role analogous to that of the smoothing parameter of an EWMA control chart: smaller values of $\eta$ concentrate the weight on the most recent observations, responding more quickly to large abrupt shifts, whereas larger values spread it more evenly over past observations, enhancing the sensitivity to small persistent shifts.
As expected, a larger $\eta$ yields a lower mean $ARL$, and thus higher performance, at small-to-moderate severity levels.
Among the values considered, $\eta = 0.98$ attains the best performance at small-to-moderate severity levels.
Since the proposed control chart is primarily intended to detect small and persistent departures from the IC state, a value of $\eta$ close to one is preferable. We fixed $\eta = 0.95$ a priori throughout. For $p = 3$, $\eta = 0.98$ would have yielded even better performance at the small-to-moderate severity levels that the method primarily targets, so that the advantage over the competitors reported in Section~\ref{sec_sim} is, if anything, conservative.

\subsection{Fraction of variance explained}
\label{supp_fve}
\begin{figure}
    \centering
    \includegraphics[width=1\linewidth]{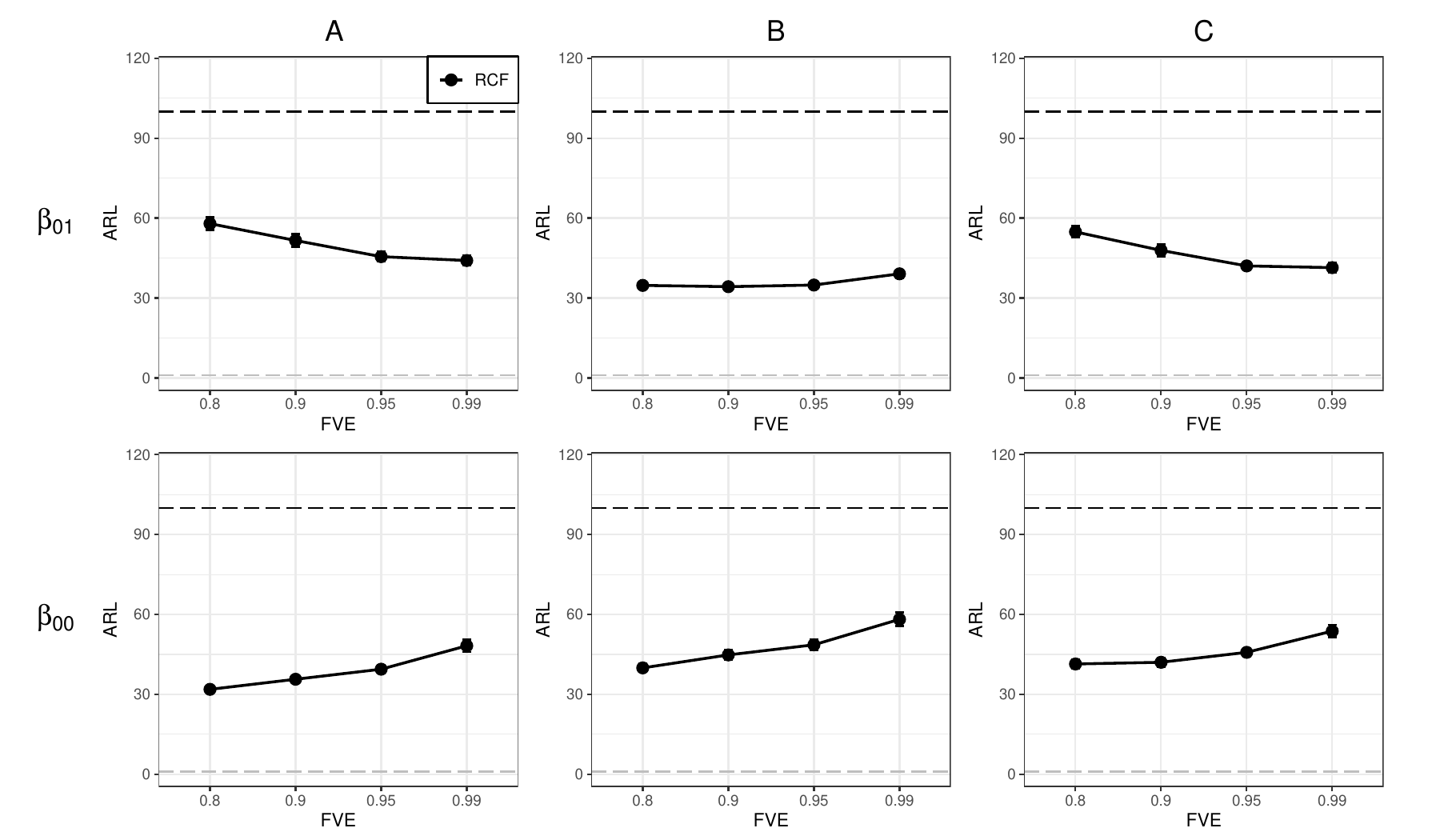}
    \caption{Mean $ARL$ over the Monte Carlo simulation runs, and its corresponding $95\%$ confidence interval achieved in Phase II by RCF, for different values of $\mathrm{FVE} = 0.8, 0.9, 0.95, 0.99$, for $p = 3$ functional covariates and $SL = 3$.}
    \label{fig:sens_fve}
\end{figure}
We vary the FVE over $\mathrm{FVE} = 0.8, 0.9, 0.95, 0.99$ at the fixed severity level $SL = 3$, holding all other parameters at their default values. Figure \ref{fig:sens_fve} reports the mean $ARL$ over the Monte Carlo simulation runs, and its corresponding $95\%$ confidence interval as a function of the FVE for each shift type (columns A, B, C) and shift location (top row, shift in the regression coefficient function $\beta_{01}(s,t)$; bottom row, shift in the intercept function $\beta_{00}(t)$).
Unlike the number of basis functions, the FVE affects the performance in a way that depends markedly on the shift type and location. This is expected, since different shifts project onto different sets of functional principal components (FPCs). When a shift concentrates on a small number of leading FPCs, a monitoring scheme that retains only those components through the $T^2$ statistic is more powerful than one that also includes further, largely unaffected components, which dilute the signal. The $SPE$ statistic captures the variation outside the retained subspace. The FVE governs this trade-off by determining how many FPCs the $T^2$ subspace retains. As noted by \cite{centofanti2025adaptive}, this implies that the parameters best for monitoring do not necessarily coincide with those best for estimation.
For a shift in the regression coefficient function $\beta_{01}(s,t)$, the behavior differs across shift types. Under shift types A and C, performance improves as FVE increases up to 0.95, beyond which it stabilizes. Under shift type B, performance is essentially the same for $\mathrm{FVE} = 0.8, 0.9, 0.95$ and deteriorates at $\mathrm{FVE} = 0.99$. Instead, for a shift in the intercept function only, the performance decreases monotonically with the FVE for all three shift types, and the mean $ARL$ increases with the number of FPCs retained. This indicates that the intercept shift projects mainly onto the leading FPCs and that retaining additional components only dilutes the signal.
Overall, no single FVE value is uniformly optimal across all shift types and locations, and practitioners cannot select the best value because the shift shape is unknown. The value $\mathrm{FVE} = 0.95$ attains the best or near-best performance for shifts in the regression coefficient function. In contrast, for shifts in the intercept function, it entails a visible, though not severe, loss relative to smaller FVE values.

\section{Plot of the FRCC applied to the case study}
\label{supp_frcc}
\begin{figure}
    \centering
    \includegraphics[width=1\linewidth]{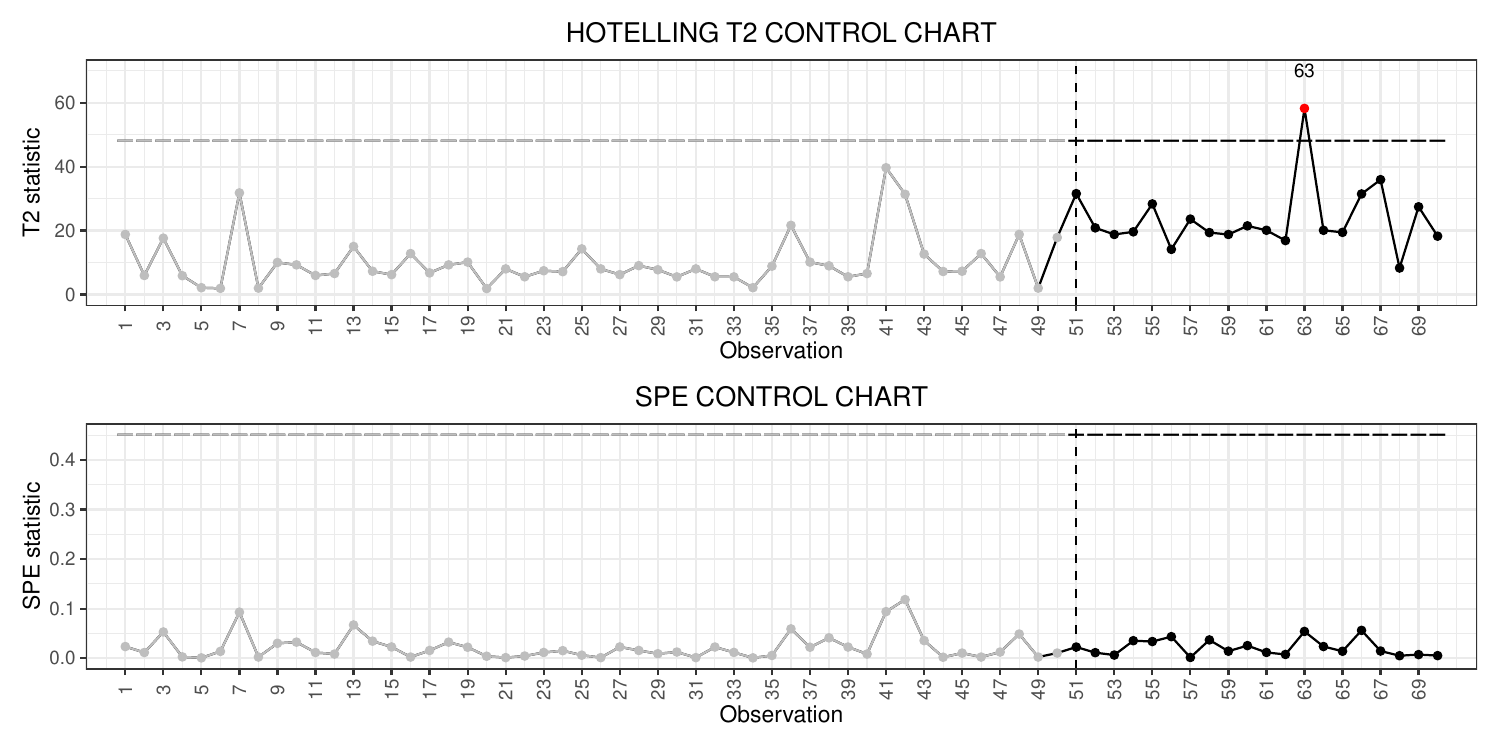}
    \caption{FRCC applied to the OC sequence data in the case study. The grey dots on the left of the dashed vertical line correspond to 50 IC observations, while the black and red dots correspond to OC observations.}
    \label{fig:frcc_cc}
\end{figure}

The FRCC is applied to the OC sequence of observations presented in the case study of Section~\ref{sec_real}. Figure \ref{fig:frcc_cc} shows the Hotelling $T^2$ control chart (top panel) and $SPE$ control chart (bottom panel), where the grey dots on the left of the dashed vertical line correspond to 50 IC observations, while the black and red dots correspond to OC observations. The $T^2$ statistic shows a small but visible upward shift over the OC region relative to the IC region, indicating that it responds, albeit minimally, to the change. However, since it monitors each observation individually, it is insensitive to the small sustained shift induced by the $\lambda$-undershoot fault, and its $T^2$ statistic exceeds the corresponding control limit only at a single isolated observation, while the $SPE$ statistic never exceeds its corresponding control limit.

\end{document}